\documentclass{aa} 

\usepackage{longtable}
\usepackage{lipsum}
\usepackage{ulem}
\usepackage{graphicx}
\usepackage{url}
\usepackage{txfonts}
\begin{document}

   \title{Activity of TRAPPIST--1 analogue stars observed with \textit{TESS}}

   \author{B\'alint Seli\inst{1,2}
          \and
          Kriszti\'an Vida\inst{1}
          \and
          Attila Mo\'or\inst{1}
          \and
          Andr\'as P\'al\inst{1}
          \and
          Katalin Ol\'ah\inst{1}}

   \institute{
   Konkoly Observatory, Research Centre for Astronomy and Earth Sciences, H-1121 Budapest, Konkoly Thege M. \'ut 15-17, Hungary
    \and
    E\"otv\"os University, Department of Astronomy, Pf. 32, 1518 Budapest, Hungary }
    
   \date{Received December 9, 2020}

 
\abstract{As more exoplanets are being discovered around ultracool dwarfs, understanding their magnetic activity -- and the implications for habitability -- is of prime importance. To find stellar flares and photometric signatures related to starspots, continuous monitoring is necessary, which can be achieved with spaceborn observatories like the Transiting Exoplanet Survey Satellite (\textit{TESS}).

We present an analysis of TRAPPIST--1 like ultracool dwarfs with \textit{TESS} full-frame image photometry from the first two years of the primary mission. A volume-limited sample up to 50\,pc is constructed consisting of 339 stars closer than $0.\!^{\rm m}5$ to TRAPPIST--1 on the \textit{Gaia} colour--magnitude diagram. The 30-min cadence \textit{TESS} light curves of 248 stars were analysed, searching for flares and rotational modulation caused by starspots. The composite flare frequency distribution of the 94 identified flares shows a power law index similar to TRAPPIST--1, and contains flares up to $E_\mathrm{\textit{TESS}} = 3 \times 10^{33}$\,erg. Rotational periods shorter than 5$^\mathrm{d}$ were determined for 42 stars, sampling the regime of fast rotators. The ages of 88 stars from the sample were estimated using kinematic information. A weak correlation between rotational period and age is observed, which is consistent with magnetic braking.}


   \maketitle
%

\section{Introduction}
\label{sec:introduction}

Low-mass, cool M-dwarfs are the prime targets when searching for exoplanets. One appealing trait of these objects is that their habitable zone is much closer to an M-dwarf host than to a solar-like star, thus detecting a possibly habitable Earth-like planet is easier. Currently there are about 4000 known exoplanets, but interestingly only very few of these were found around ultracool dwarfs (UCDs, late M dwarfs and brown dwarfs). One of these is Proxima Centauri (M5.5 V) --  the closest star to the Sun -- that hosts an Earth-mass planet in its habitable zone \citep{proxima-planet}. The other one is Teegarden's star (2MASS J02530084+1652532; M7V), where \cite{teegarden-carmenes} recently reported two Earth-mass planets, that are among the lowest-mass planets discovered so far. The third example is TRAPPIST-1 (2MASS J23062928-0502285; \textit{Gaia} DR2 2635476908753563008; M8V), that is known to host seven transiting exoplanets, three of which have equilibrium temperatures that makes the existence of liquid water on their surface possible \citep{trappist1}. However, due to the low luminosity of these objects in the optical regime, obtaining high signal-to-noise photometric or spectroscopic observation is challenging, hence the low number of currently known exoplanets around UCDs.

The habitability of the planets hosted by UCDs is in the focus of intense debates, as a large fraction of them is magnetically active, producing e.g. frequent flares.  These can damage the atmospheres of the orbiting planets by changing the atmospheric composition or completely eroding the atmosphere over time \citep[see][and references therein]{t1-flares, t1-rachael, proxima-flares, evryscope}. On the other hand these events can be the source of the UV radiation needed for abiogenesis \citep{flares-prebiotic}. On our Sun a large fraction of the flares are accompanied by coronal mass ejections (CMEs), that also can have deleterious effects on planetary atmospheres. However, observations suggest that while the detected CMEs are more frequent on the later-type, more active UCDs, most of these events are unsuccessful eruptions, with velocities below the escape velocity \citep{v374peg, CMEs}. This is further confirmed by magnetohydrodynamic simulations, that suggests that only the largest eruptions might be able to escape from the strong magnetic field of these stars \citep{CME-MHD}.

To understand these systems, long-term observations would be optimal. However, in this paper we chose a different approach: instead of observing a single target for an extended period of time, we aim to gain statistical knowledge of the flaring activity based on the photometry of many similar stars. To this end, we choose TRAPPIST--1 as the main target, and compile a list of similar objects we call \textit{TRAPPIST--1 analogues}. As a measure of similarity we simply use the distance between stars on the \textit{Gaia} colour--magnitude diagram, and adopt a lower limit on parallax to have a volume-limited sample. As for the photometric observation, we use data provided by the {\it Transiting Exoplanet Survey Satellite} mission \citep[\textit{TESS};][]{ricker2015}. \textit{TESS} performed its primary mission between July 2018 and July 2020, covering the majority of the sky divided into observation runs called {\it sectors}. One sector covers $24\times 96$ degrees in a nearly continuous field-of-view (FoV), assembled from the $24\times 24$ degree FoVs of four individual identical cameras lined up to provide the final geometry. The two years of the primary mission were divided into a Southern Survey (1st year) and Northern Survey (2nd year), where each of the corresponding $2\times 13$ sectors were observed for approximately one sidereal month ($\sim27.3^\mathrm{d}$). The sectors were arranged in a way to provide overlaps and therefore nearly continuous observations for a year at the vicinity of the ecliptic poles\footnote{\url{https://tess.mit.edu/observations/}}. \textit{TESS} images are available for the entire FoV with 30-min cadence, while a limited number of stars are observed with 2-min cadence. To make our analysis homogeneous, we only use the 30-min full-frame images (FFIs), which are available for $\sim 3/4$ of virtually any spatially homogeneous stellar sample (in the two years of the \textit{TESS} primary mission, which avoids the ecliptic). Our primary objective is to detect the rare high energy flares, and many studies have shown that it is possible with this relatively long cadence (see e.g. \citealt{lc_flares1, lc_flares2}). This cadence is also sufficient for the detection of photometric rotational period of UCDs in the order of a few days.

The structure of the paper is as follows: we define the TRAPPIST--1 analogue sample in Sect.~\ref{sec:sample}, then in Sect.~\ref{sec:methods} the analysis of the \textit{TESS} light curves is presented, including period search and flare identification. In this section we also look for H$\alpha$ emission in publicly available optical spectra, and compile the approximate ages of stars in the sample. We then discuss the implications of our findings in Sect.~\ref{sec:discussion}, regarding the period distribution, flare rate and possible correlations. Finally the main results are summarised in Sect.~\ref{sec:conclusions}.

\section{Sample selection}
\label{sec:sample}

\begin{figure}
    \centering
    \includegraphics[width=\columnwidth]{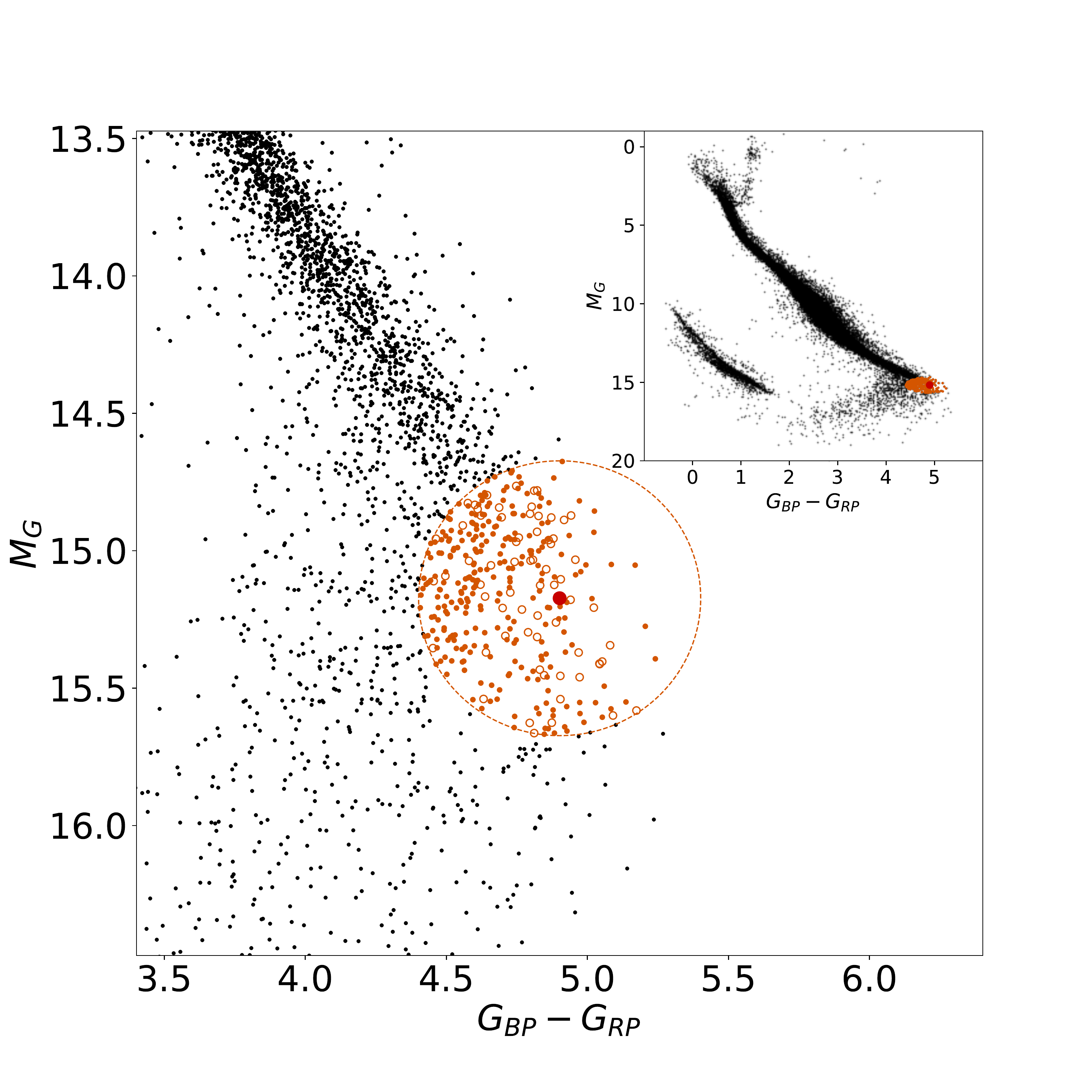}
    \caption{Selection criteria on the \textit{Gaia} colour--magnitude diagram. Red point shows the position of TRAPPIST--1, orange points represent the final TRAPPIST--1 analogue sample. Empty orange circles indicate stars that were not initially included due to the \textit{Gaia} quality cuts, but were later added from existing UCD catalogues.}
    \label{fig:hrd}
\end{figure}

Since TRAPPIST--1 lies near the ecliptic, \textit{TESS} did not observe it in the first two years of the primary mission, and the next opportunity will be during the fourth year, when Sectors 42--46 will be centered on the ecliptic. So to study UCDs like TRAPPIST--1, we selected similar stars (henceforth TRAPPIST--1 analogues) based on simple photometric criteria. The selected stars need to be closer than $0.\!^{\rm m}5$ to TRAPPIST--1 on the \textit{Gaia} $(G_{BP}-G_{RP})$ -- $M_G$ diagram (see Fig.~\ref{fig:hrd}). This results in a circle around TRAPPIST--1, which has the following parameters: $G_{BP}-G_{RP} = 4.\!^{\rm m}901 \pm 0.\!^{\rm m}049$, $M_G = 15.\!^{\rm m}1728 \pm 0.\!^{\rm m}0035$. To get a volume limited sample for homogeneous analysis, only stars closer than 50\,pc were considered. Limiting our search to nearby stars is also practical so our sample will contain only targets within the reach of \textit{TESS}. According to TICgen\footnote{ \url{https://github.com/tessgi/ticgen}}, TRAPPIST--1 at 50\,pc would be $\sim17^{\rm m}$ in the \textit{TESS} bandpass with photometric error around $0.\!^{\rm m}07$ on 30-min exposures, enough to detect the highest energy flares. This value agrees with the pure \textit{Gaia} DR2-based estimation of $17.\!^{\rm m}10$. Such an estimation is rather accurate due to the nearly perfect overlap of the bandpasses of the \textit{TESS} cameras and the \textit{Gaia} $G_\mathrm{RP}$ colour \citep[see Fig.~3 in][]{jordi2010}.

\begin{table*}
\caption{SQL query used to select the sources from the \textit{Gaia} DR2 catalogue using the services provided by the \textit{Gaia} TAP server.}
\label{table:sql}
\small
\begin{center}
\begin{verbatim}
SELECT *
  FROM gaiadr2.gaia_source
WHERE parallax > 20
  AND parallax_over_error > 10
  AND phot_g_mean_flux_over_error > 50
  AND phot_rp_mean_flux_over_error > 20
  AND visibility_periods_used > 8
  AND astrometric_chi2_al/(astrometric_n_good_obs_al-5) < 1.44*greatest(1,exp(-0.4*(phot_g_mean_mag-19.5)))
\end{verbatim}
\end{center}
\end{table*}

We used a custom query (Table~\ref{table:sql}) on the \textit{Gaia} TAP (Table Access Protocol) server\footnote{\url{http://gea.esac.esa.int/tap-server/tap}} to obtain the set of stars with reliable observations within 50\,pc. The flux error limits employed in this query were based on the work of \cite{gaia-filter}. The last line of the query corresponds to the Renormalised Unit Weight Error (RUWE), retaining sources with well-behaved astrometric solutions. Since the stars of interest are in the solar neighbourhood, reddening and interstellar extinction were not taken into account.

After applying our photometric criteria on this list 271 stars remained, with 144 also present in the UCD catalogue of \cite{gaia-filter}. To complement this list with known UCDs (since e.g. TRAPPIST--1 itself was excluded by RUWE), we also added stars from the \textit{List of M6-M9 Dwarfs}\footnote{\url{https://jgagneastro.com/list-of-m6-m9-dwarfs/}} curated by J. Gagn\'e, disregarding the quality criteria used in the previous \textit{Gaia} query. Also we included 36 new stars from the volume-limited \textit{Ultracool SpeXtroscopic Survey} of \cite{SpeX} in the same manner. All the added stars are within 50\,pc and inside the $0.\!^{\rm m}5$ circle on the \textit{Gaia} colour--magnitude diagram, their exclusion from the original sample was due to either \mbox{\textit{visibility\_periods\_used}} $\leq 8$ or RUWE. The final sample consists of 339 targets, summarised in Table \ref{table:targets}. Propagating the \textit{Gaia} uncertainties of TRAPPIST--1 to the selection, the scatter in the sample size is $\pm39$. As a sanity check we queried the Simbad database with the \textit{Gaia} DR2 identifiers, and $\sim90\%$ of the valid spectral types were indeed between M7 and M9. We note that while the \textit{Gaia} EDR3 is now available, it could add only a few UCD candidates to the sample. According to \cite{gaia_edr3}, the number of new UCDs up to 100\,pc is 1016 objects, most of them in the faint regime that can not be reliably observed with \textit{TESS}.

\begin{figure}
    \centering
    \includegraphics[width=\columnwidth]{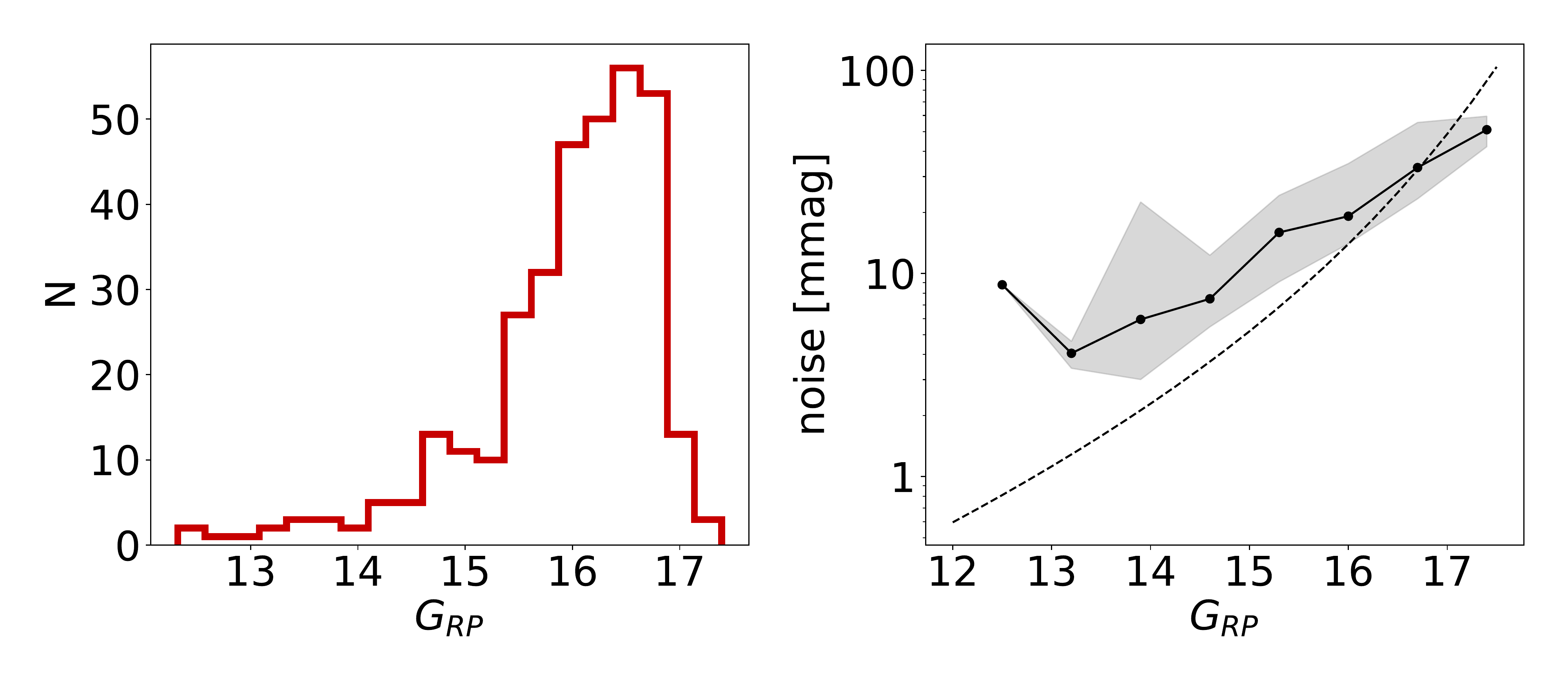}
    \caption{\textit{Left:} Brightness distribution of the final sample. \textit{Right:} Noise properties of the sample. The black curve shows the photometric scatter measured on the final processed light curves. The region between the 16th and 84th percentiles is shown in grey. The dashed line is the predicted noise value from TICgen.}
    \label{fig:hist}
\end{figure}

\begin{figure*}
    \centering
    \includegraphics[width=\textwidth]{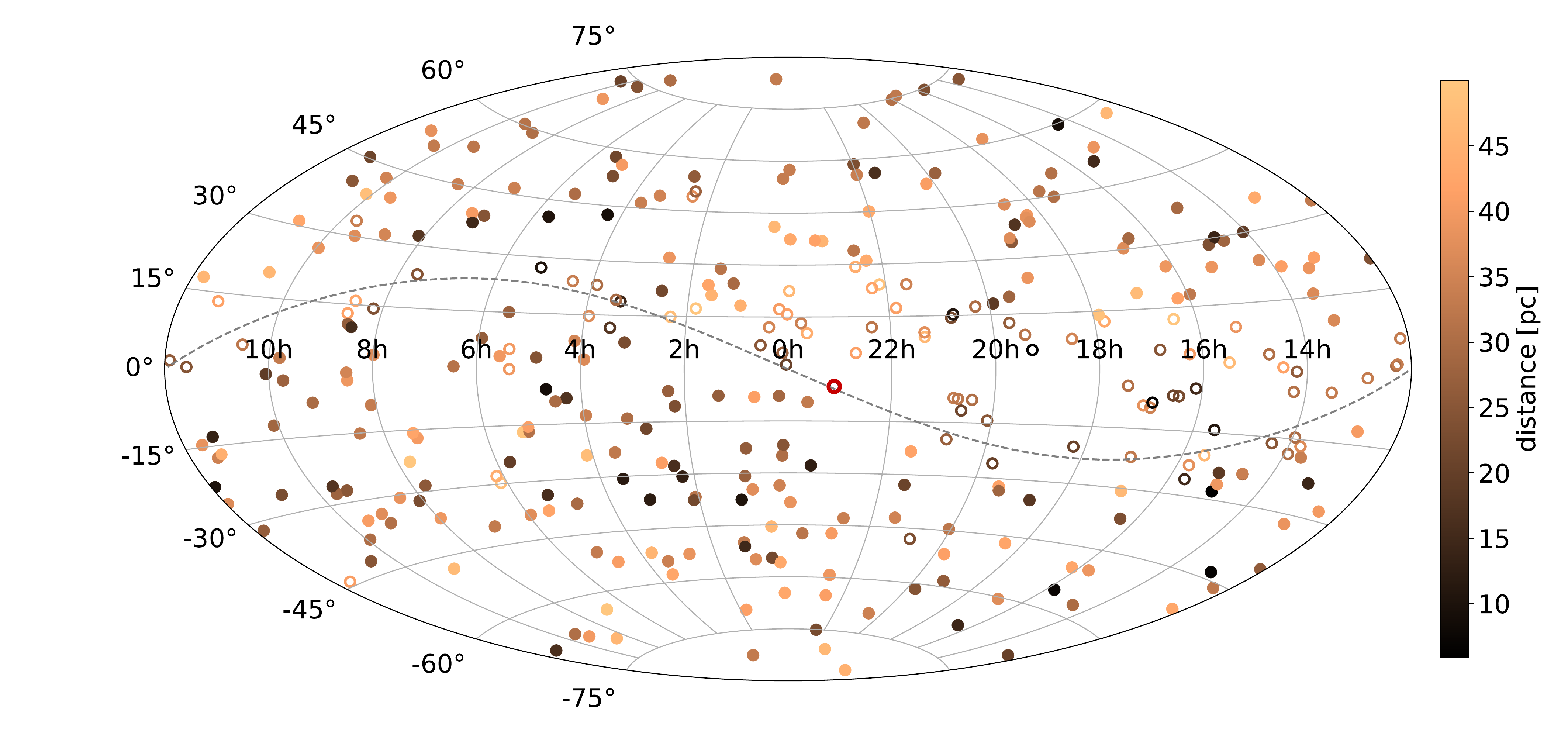}
    \caption{Position of the selected stars on the sky with equatorial coordinates, colour coded with distance. Stars plotted with filled points have been observed by \textit{TESS} up to Sector 26. Red circle represents TRAPPIST--1. The ecliptic is shown with dashed line for reference.}
    \label{fig:map}
\end{figure*}

The brightness distribution of our sample can be seen in Fig.~\ref{fig:hist}. Since the \textit{Gaia} $G_\mathrm{RP}$ bandpass is similar to the \textit{TESS} bandpass, we expect most targets to be brighter than $T=17^{\rm m}$. Fig.~\ref{fig:map} shows the position of the targets with equal-area Aitoff projection. Open circles indicate stars that were not observed in the \textit{TESS} primary mission, mostly due to the exclusion of the ecliptic.

325 stars from the sample had entry in the \textit{TESS} Input Catalog \citep[TICv8,][]{tic}, with the following median parameters: $\log{g} = 5.27 \pm 0.01$, $M = 0.092 \pm 0.003 M_{\sun}$, $L = 0.0007 \pm 0.0004 L_{\sun}$, $R = 0.116 \pm 0.003 R_{\sun}$. While there are no effective temperature measurements below 2700\,K in TICv8, there is a generally good agreement of the other parameters compared to \cite{t1_reanalysis}. 70 stars from the sample have temperature measurement in \textit{Gaia} DR2, with $3318 \pm 23$\, K median value, while TRAPPIST--1 itself has $3352_{-62}^{+104}$\,K. However, the \textit{Gaia} temperatures are not reliable in this regime, since the algorithm was trained on stars hotter than 3000\,K, and there is a degeneracy between temperature and extinction/reddening \citep{gaia_dr2}.

\begin{table*}
\caption{The TRAPPIST--1 analogue sample. The \textit{TESS} sector column shows only sectors where the target was successfully observed (up to Sector 26). For the description of the last three columns see Sect.~\ref{sec:halpha} and \ref{sec:age}. The full table will be available online.}\label{table:targets}
\centering
\tiny
\begin{tabular}{ccccccccc}
\hline \hline
Gaia DR2 & TIC & $G_\mathrm{BP}$ [mag] & $G_\mathrm{RP}$ [mag] & distance [pc] & \textit{TESS} sector & H$\alpha$ EW [\AA] & age [Gyr] & comment on age\\
\hline
3533892921879936 & 377062703 & 20.56 & 15.73 & $23.75 \pm 0.09$ & 4  & - & - & - \\
16386404041502592 & 416670281 & 20.38 & 15.29 & $17.89 \pm 0.05$ & - & $-5.4 \pm 1.8$ & - & - \\
39612036694609152 & 242960146 & 21.36 & 16.70 & $36.80 \pm 0.33$ & - & $9.3 \pm 3.6$ & $0.8 \pm 0.1$ & HYA candidate (99.5\%) \\
56252256123908096 & 5579232 & 19.21 & 14.57 & $14.65 \pm 0.03$ & - & $20.8 \pm 8.4$ & $3.1 \pm 3.4$ & kinematics \\
57744155965790720 & 456938550 & 20.80 & 16.01 & $29.31 \pm 0.36$ & - & - & - & - \\
\dots & \dots & \dots & \dots & \dots & \dots & \dots & \dots  & \dots \\
\hline
\end{tabular}
\end{table*}

\section{Methods}
\label{sec:methods}
\subsection{TESS full-frame image photometry}

We used all 26 sectors from the \textit{TESS} primary mission, which covered $\sim$75\% of the sky. 248 stars (73\% of the 339 objects) from the sample were observed in at least one sector, resulting in 370 individual light curves. The FFIs were acquired with 30-min cadence during the primary mission and the calibrated imaging data have been downloaded from the MAST bulk download portal\footnote{\url{http://archive.stsci.edu/tess/bulk_downloads.html}}. Starting from Sector 27, the TESS FFIs are downlinked with 10-min cadence, providing better time resolution at the expense of larger photometric noise. To make the analysis homogeneous we only use Sectors 1--26 here, but the data from the extended mission will provide an interesting comparison that might be addressed in a future work.

For the extraction of light curves from a series of FFI stamps, we used a flux extraction method based on convolution-aided differential photometry named \textsc{qdlp-extract} which has been implemented atop the tasks of the \textsc{FITSH} package \citep{fitsh}. We used apertures with 1.5\,pixel radius and an annulus from 5 to 10\,pixels from the target as background to have an estimate on the fluxes and their respective uncertainties. 
Since most of our targets are faint and lie in dense regions on the sky, the raw \textit{TESS} light curves often contain astrophysical signals that are originated from brighter stars. To mitigate this contamination, we employed a Principal Component Analysis (PCA) based technique to remove both instrumental and astrophysical systematics from the light curves, similar to the method used by \cite{k2_pca}. There are also cotrending basis vectors provided by TESS (for each sector and each camera), but we chose to do PCA locally, using light curves extracted with the same method as the target, to have a homogeneous basis. Around each of our targets we selected multiple stars (up to 20), and extracted their light curves as well. Then by running  PCA on these datasets, the light curve of the target can be rebuilt from several PCA components containing the variation of the nearby stars. Finally, by subtracting the PCA reconstruction from the raw light curve we can in theory remove the systematics.

The median proper motion of the sample is $0.\!^{\prime\prime}4/{\rm yr}$, which is small compared to the $21^{\prime\prime}$ pixel size of \textit{TESS}. Nevertheless, we corrected the coordinates of the targets for the $\sim$4 year difference between the \textit{Gaia} and \textit{TESS} observations. For only 6 targets was this correction larger than one \textit{TESS} pixel.

\begin{figure}
    \centering
    \includegraphics[width=\columnwidth]{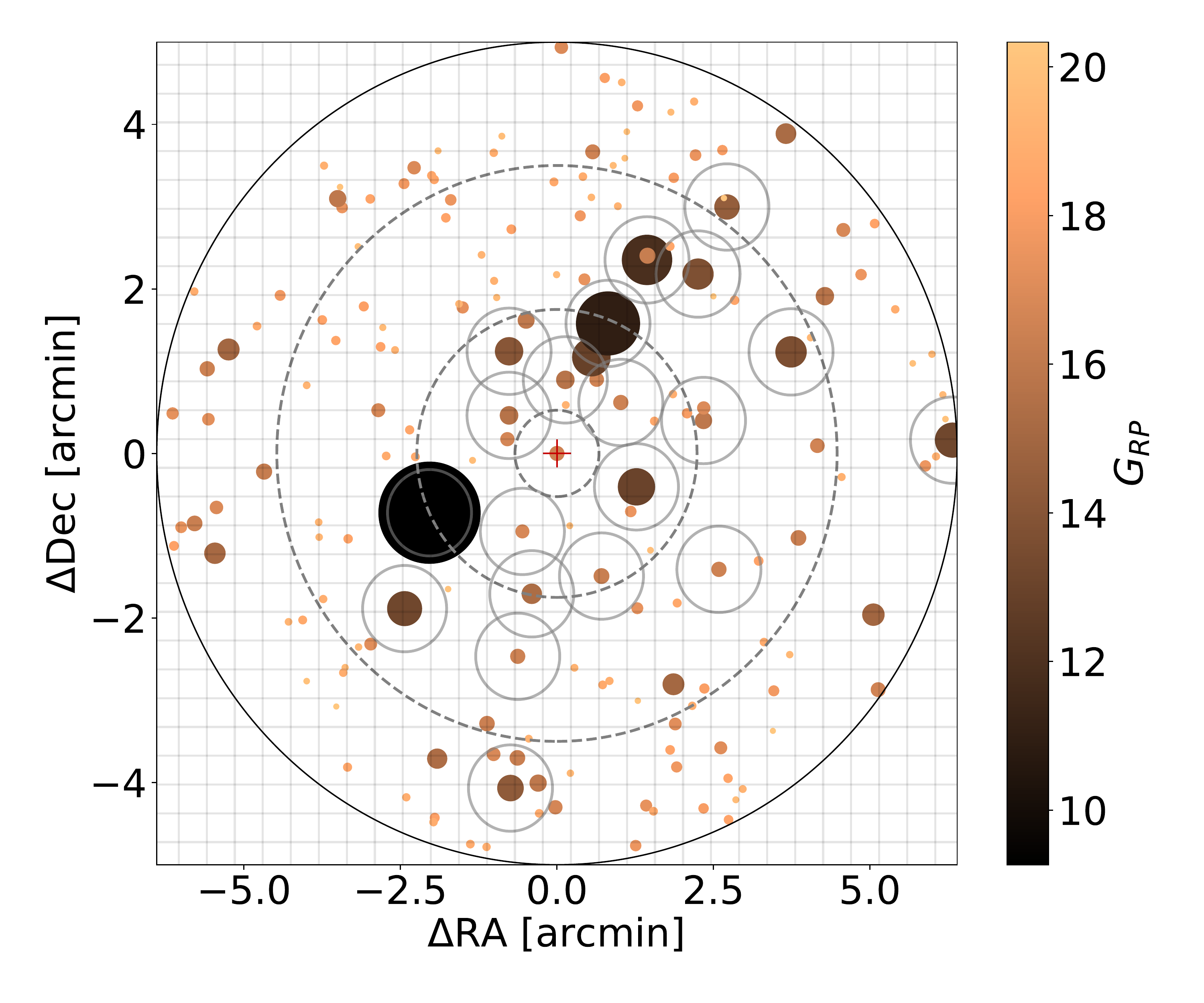}
    \caption{The field of view of a selected target (Gaia DR2 2883680659313632896) from \textit{Gaia}, with 173 stars brighter than $20^\mathrm{m}$. Red cross marks the target, dashed circles show the aperture and annulus used for the \textit{TESS} photometry, and solid circles show the stars selected for PCA. The grid in the background illustrates the pixel scale of \textit{TESS}.}
    \label{fig:grid}
\end{figure}

\begin{figure}
    \centering
    \includegraphics[width=0.95\columnwidth]{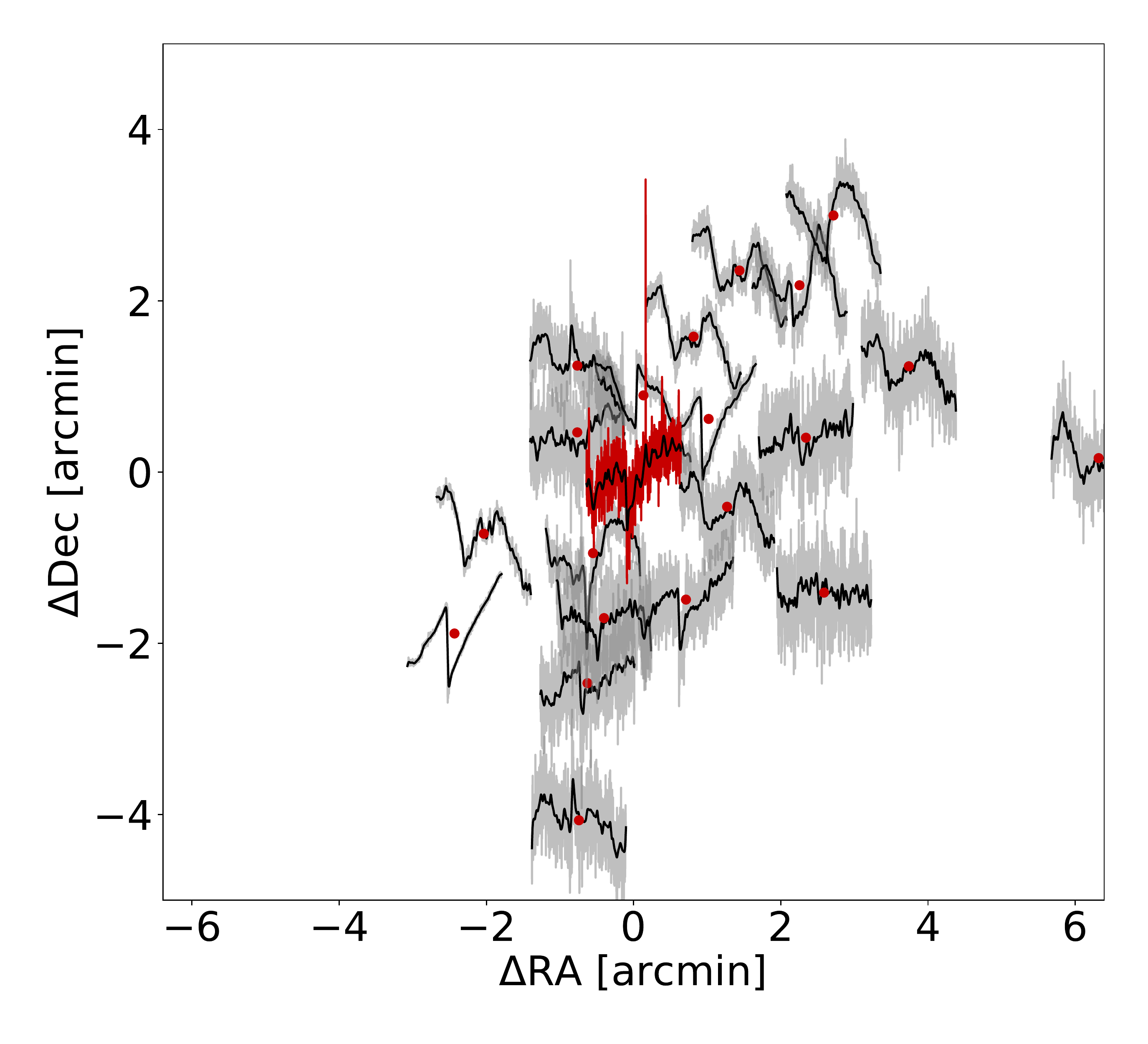}
    \caption{Extracted light curves from the circular apertures of Fig.~\ref{fig:grid}, with red dots indicating the celestial position of the sources. Some common trends can be identified, while the large flare seen on the target (red curve) does not seem to appear anywhere else.}
    \label{fig:light_curves_for_pca}
\end{figure}

To find the possible contaminating stars, we queried \textit{Gaia} DR2 around our targets with $5^{\prime}$ radius. We corrected the position of all stars for proper motion (except for stars with less than 5 astrometric parameters solved, i.e., with \mbox{\textit{astrometric\_params\_solved}} $\ne 31$). We selected the stars closer than $3^{\prime}$ and at most $0.\!^{\rm m}5$ fainter than the target ($G_\mathrm{RP} < G_\mathrm{RP,target} + 0.\!^{\rm m}5$). We also added stars up to $5^{\prime}$ if they are at least $2^{\rm m}$ brighter than the target. From this list, the brightest 20 stars were kept, leaving out objects closer than $0.\!^{\prime}5$ ($\sim1.5$ \textit{TESS} pixel) to each other, also leaving out stars closer than $0.\!^{\prime}5$ to the target itself. An example of such a selection can be seen in Fig.~\ref{fig:grid}, while Fig.~\ref{fig:light_curves_for_pca} shows the extracted light curves for the selected stars around the same target.

\begin{figure}
    \centering
    \includegraphics[width=\columnwidth]{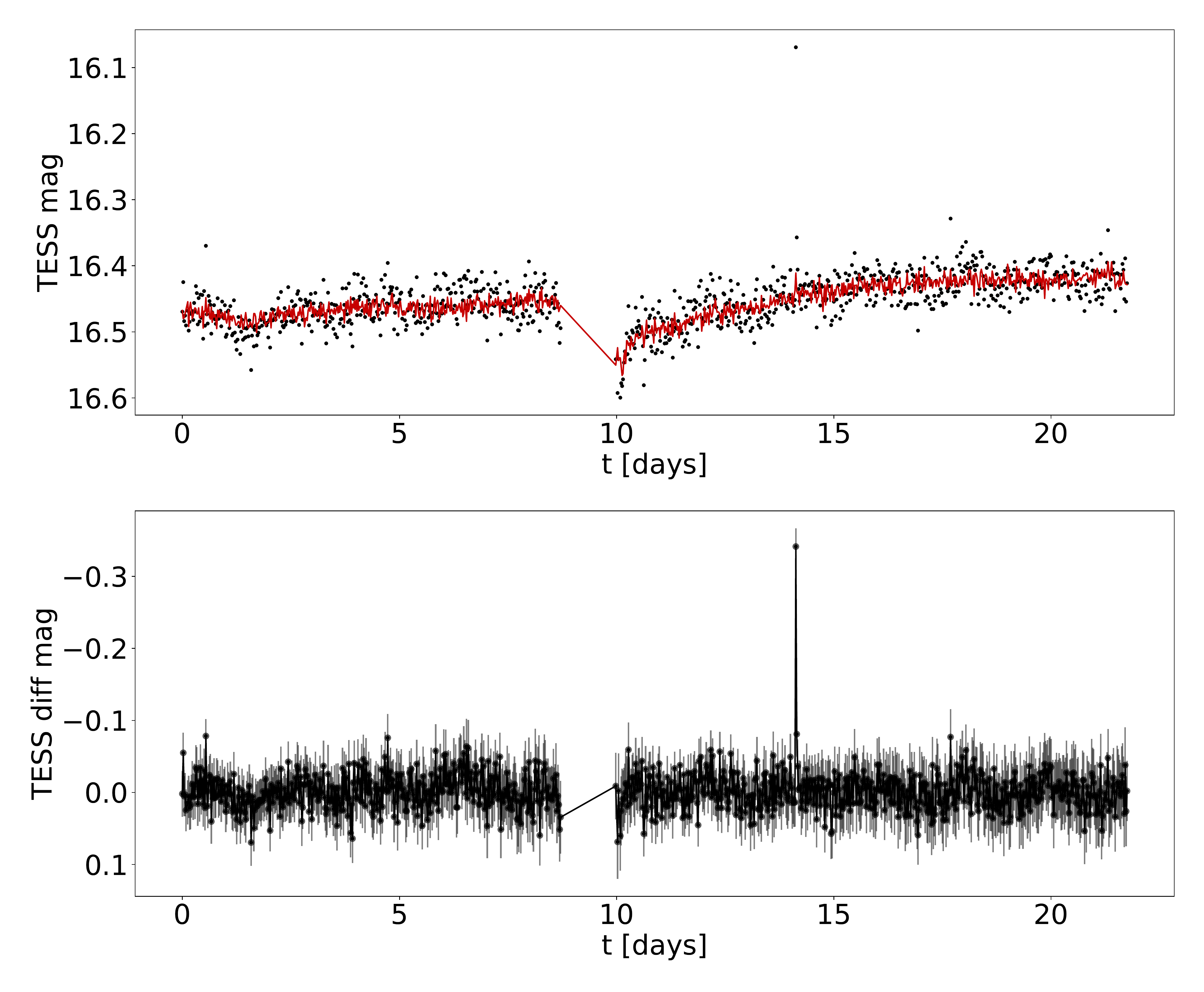}
    \caption{PCA reconstruction of the light curve from Fig.~\ref{fig:light_curves_for_pca}. Black points on the upper panel show the light curve created by \textsc{qdlp\_extract}, red curve shows the PCA reconstruction. The corrected light curve can be seen on the lower panel, with a dominant flare and periodic variation.}
    \label{fig:pca_reconstruction}
\end{figure}

As a post-processing of the light curves, we removed bad photometric points with quality flags masked with the bitmask 10101111 \citep[manual exclude, reaction wheel desaturation, Earth pointing, coarse pointing, safe mode or attitude tweak, see also][]{tenenbaum2018}, and also 1\% of the points with the largest error bars on the given light curve. We also removed NaN values that were possibly due to zero measured flux or instrumental errors. Then the light curves of the nearby stars used for PCA were interpolated to the time values of the main target, so that all datasets contain the same number of points. We also tried the same approach with flux instead of magnitude, but the results were essentially the same. A light curve containing periodic variation and a flare event can be seen in Fig.~\ref{fig:pca_reconstruction}.

\subsection{Period search}

Ultracool dwarfs often have rotational periods less than a few days, therefore even one $27^\mathrm{d}$ long \textit{TESS} sector could be enough to detect the rotational modulation caused by starspots. To search for rotational periods, we inspected the Lomb--Scargle periodogram \citep{Lomb, Scargle} of all stars in the observed sample (248 stars out of 339, with 370 light curves). To make the analysis homogeneous, the periodograms were plotted for each sector separately, and only periods shorter than $5^\mathrm{d}$ were considered (similar to \citealt{medina_ucd_15pc} or \citealt{tess_sector1_max}). Several targets showed Lomb--Scargle peaks longer than $5^\mathrm{d}$, but the strong contamination and short temporal baseline would make these detections ambiguous.

After clipping the outliers from the light curve with $3\sigma$ threshold, the Lomb--Scargle periodogram was plotted, and the 5 largest peaks were identified between 1.5\,hours and $5^\mathrm{d}$. We folded the light curves with these trial periods, and inspected the results manually. Since all the stars with detected period showed simple sinusoidal variation, only one Fourier term was used for the analysis.

To assess the significance of the periodogram peaks, the False Alarm Probability (FAP) was calculated. The FAP quantifies the probability that a peak with the given height is observed from Gaussian noise alone, purely by chance. We note that a FAP value close to zero does not mean that the given period is the correct one. Only peaks with FAP less than 1\% were considered.

To calculate the uncertainty of the peak position, we took the best fitting sinusoidal model, and created 1000 realisations adding Gaussian noise with the scatter of the residual light curve. The period analysis was repeated, and the position of the largest peak was saved. The uncertainty of the period was then calculated as the standard deviation of these periods. We note that this uncertainty generally scales with the value of the period itself, increasing for longer periods.

Most of the stars with detected period show simple sinusoidal light curves, possibly due to the relatively long cadence compared to the short rotational periods. If these stars would show complex rotational modulation \citep[e.g. as in][]{complex_m_dwarf_modulation}, the 30-min integration time would blur the sharp features. The distribution of the detected rotational periods is plotted in Fig.~\ref{fig:kde}. We return to its discussion in Sect.~\ref{sec:discussion}.

\begin{figure}
    \centering
    \includegraphics[width=\columnwidth]{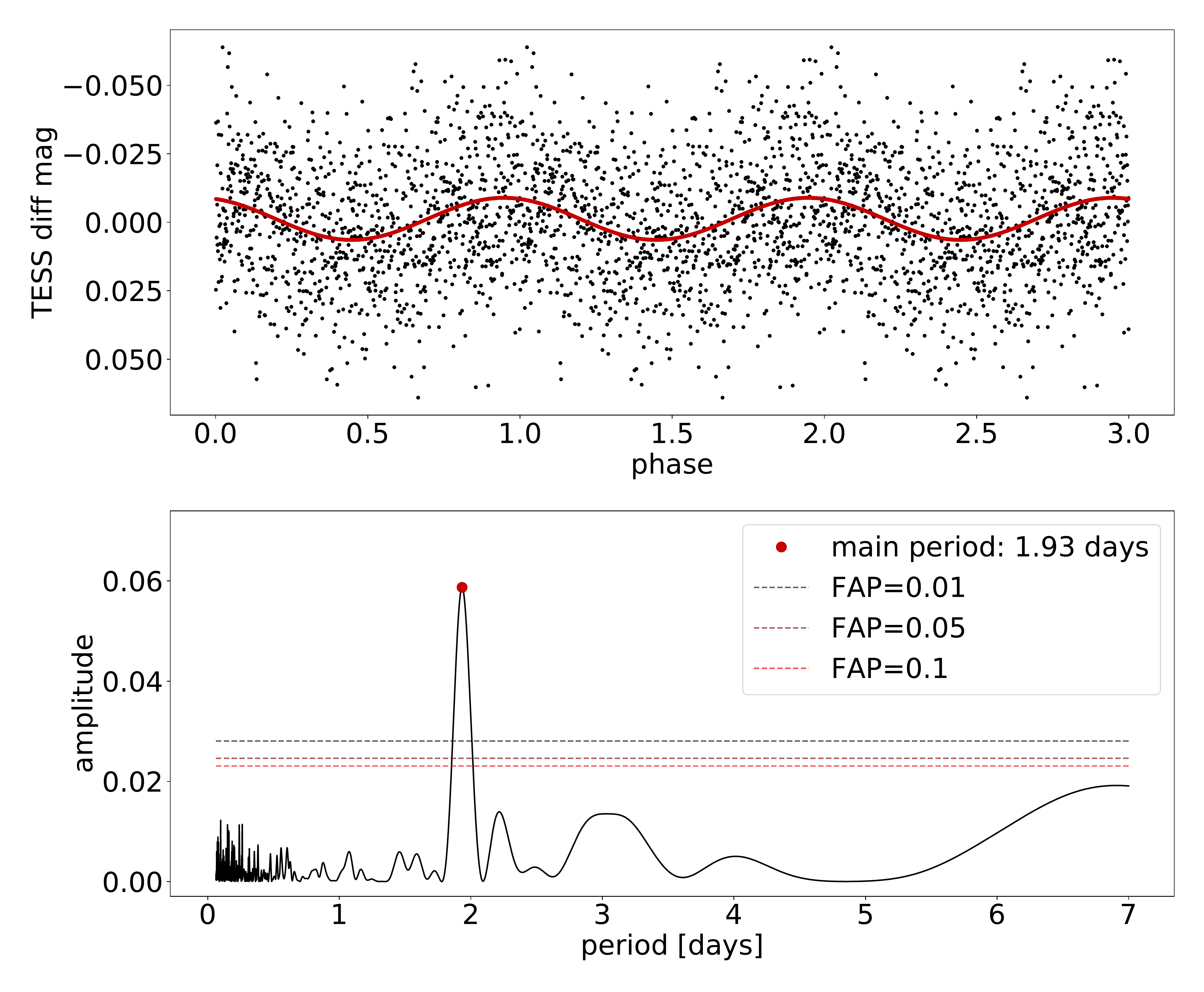}
    \caption{Period analysis for the corrected light curve from Fig.~\ref{fig:pca_reconstruction}. Upper panel shows the light curve phase folded with the 1.93$^\mathrm{d}$ period, which can be identified on the Lomb--Scargle periodogram below.}
    \label{fig:LS_periodogram}
\end{figure}

\begin{figure}
    \centering
    \includegraphics[width=\columnwidth]{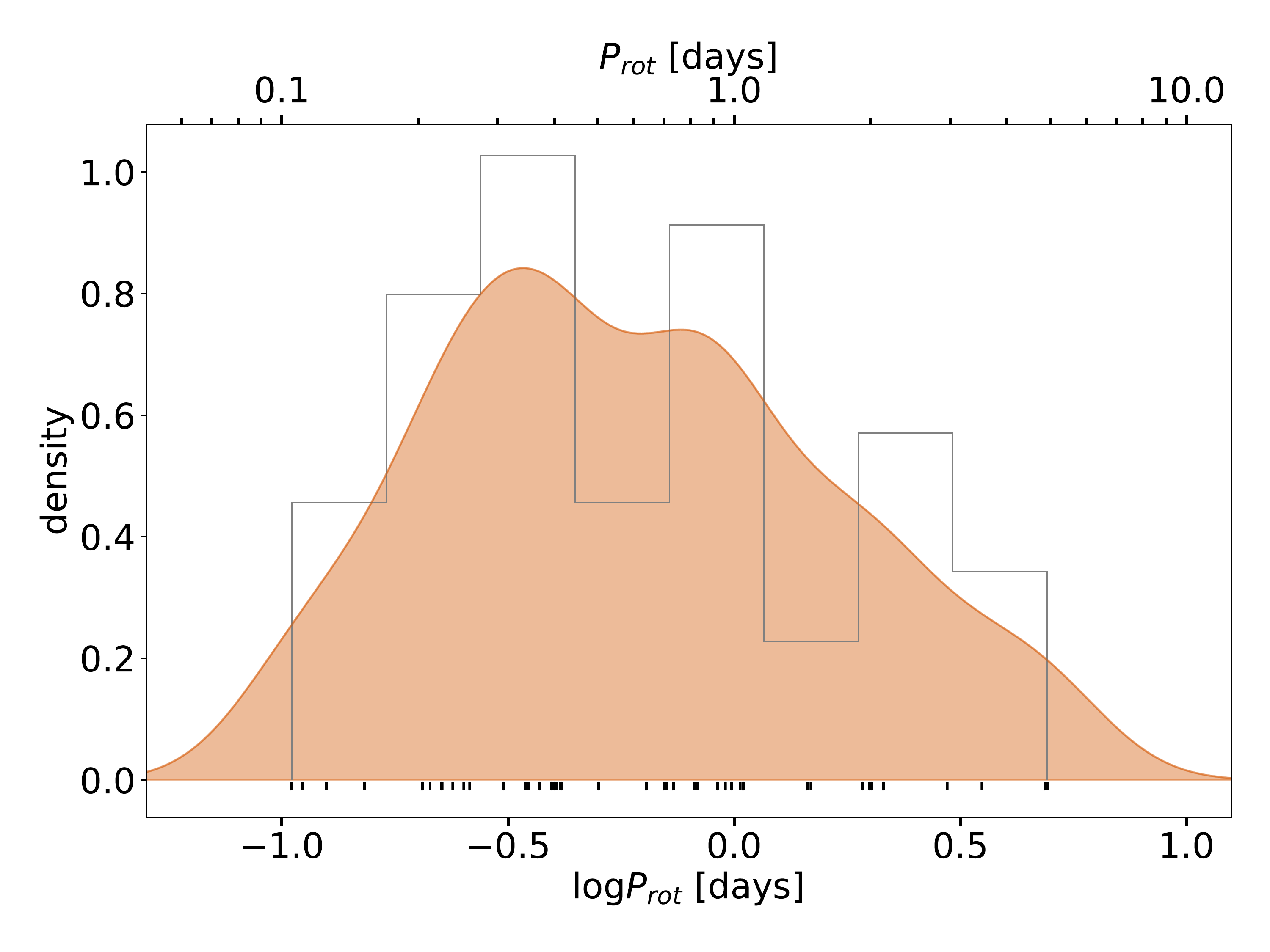}
    \caption{Rotational period distribution of the sample with \textit{TESS}. The one-dimensional kernel density estimation plotted with orange is calculated with a Gaussian kernel with bandwidth of 0.15. The black ticks show the individual period values. We note that due to the 27$^\mathrm{d}$ baseline of \textit{TESS} sectors, only periods below 5$^\mathrm{d}$ were kept.}
    \label{fig:kde}
\end{figure}

\subsection{Flares}
\label{sec:flares}

Flares appear as sudden brightening of the stellar atmosphere that occur when the magnetic field lines reconnect, and a portion of magnetic energy is released. Their usual timescale is minutes to hours. Even though white-light flares are rarely observed on the Sun, they are quite common on later-type stars (see e.g. \citealt{sun_white_light_flare}). Apart from the rotational modulation caused by starspots, flares are the most easily observable manifestation of magnetic activity in optical light curves. So to characterize the activity of TRAPPIST--1 analogues, we searched for flares on the 30-min \textit{TESS} light curves.

Due to the high noise level of the light curves, we chose to identify flares manually. We selected events with at least two consecutive outlying points (with at least one point exceeding twice the local scatter), resulting in 94 detections. These flares were found on 21 stars, all with detected rotational period, which means that 50\% of the fast rotators in our sample show flares (similar to the $\sim 60$\% found by \citealt{tess_sector1_max}).

To characterise the flares, we calculated the equivalent duration (ED, \citealt{flare_ed}), i.e. the integrated area under the flare curve (the time needed for the quiescent star to radiate the same amount of energy that was released during the flare event). The total energy output of an event can be calculated by multiplying this value with the quiescent luminosity of the star. 
We converted the light curves to flux, and ran the full detrending procedure, which involves normalising by the median value, then subtracting the PCA reconstruction, resulting in dimensionless flux centered on zero. We then iteratively $\sigma$-clipped points with lower and upper rejection threshold of $3\sigma$ and $2\sigma$, respectively, which effectively removed the flares. We then smoothed the remaining dataset with a LOWESS filter (Locally Weighted Linear Regression, \citealt{lowess}) with Gaussian kernel width of 0.06$^\mathrm{d}$. This smoothed dataset was then interpolated onto the time frame of the original light curve, and subtracted from it. This step removed the variation induced by starspots. We calculated the ED on these light curves employing two different approaches. First, we simply integrated the area of the flares using the trapezoidal rule. To estimate the uncertainty of this method, we re-sampled the light curve using the flux values and error bars as the mean and scatter of a Gaussian distribution. We generated 1000 such realisation of the light curve, calculated the ED again, and took the standard deviation.

As a second approach, we fitted the flares with the single-peaked empirical template of \cite{Davenport_template}, parameterized by the $t_\mathrm{peak}$ time of the peak, $A$ amplitude and $t_{1/2}$ width of the flare. After transforming the measured time $t$ to $t' = \frac{t - t_\mathrm{peak}}{t_{1/2}}$, the flare template function is given by:
\begin{equation}
    \begin{cases} 0 &\mbox{if } t' < -1 \\
A \cdot (1 + 1.941  t' - 0.175  t'^2 - 2.246  t'^3 - 1.125 t'^4) & \mbox{if } -1 \leq t' < 0 \\
A \cdot (0.6890 e^{-1.600 t'} + 0.303 e^{-0.2783 t'}) & \mbox{if } 0 \leq t' 
\end{cases}
\end{equation}
To fit the observations we used the Markov Chain Monte Carlo (MCMC) implementation of \textsc{pymc} \citep{pymc} with uniform prior around an approximate $t_\mathrm{guess}$ on $t_\mathrm{peak}$, and exponential priors on $A$ and $t_{1/2}$, where $\lambda$ is the parameter of the exponential distribution:

\begin{equation}
    t_\mathrm{peak} \sim \mathcal{U}(t_{guess} - 1^\mathrm{d}, t_\mathrm{guess} + 1^\mathrm{d})
\end{equation}
\begin{equation}
    A \sim \mathrm{Exp}(1 / \lambda = 0.4)
\end{equation}
\begin{equation}
    t_{1/2} \sim \mathrm{Exp}(1 / \lambda = 0.2^\mathrm{ d})
\end{equation}

\begin{figure}
    \centering
    \includegraphics[width=\columnwidth]{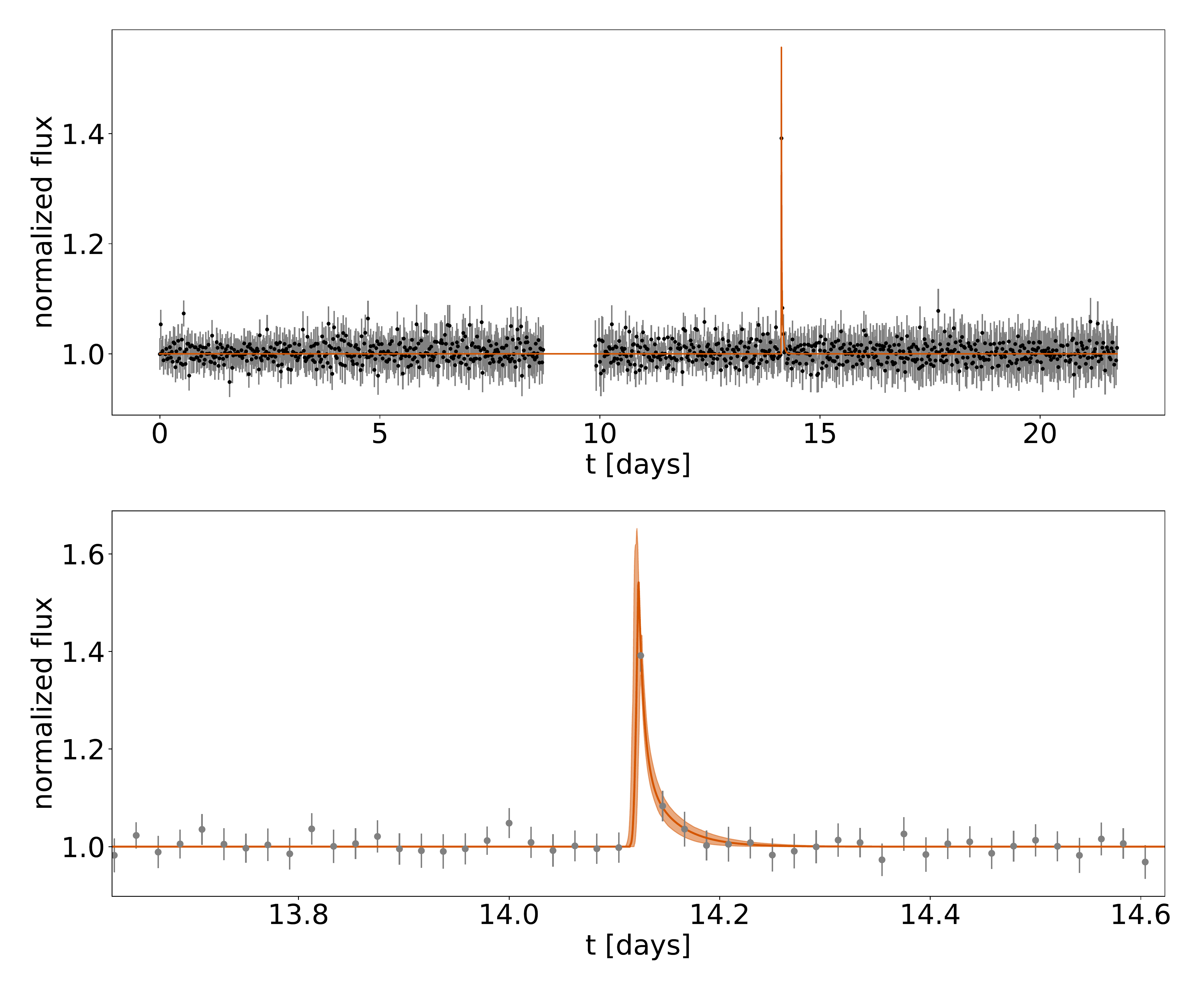}
    \caption{A flare event fitted with the analytical template following the example of Fig.~\ref{fig:pca_reconstruction}. The summed ED is $15.4 \pm 2.9$\,min, while the fit gives $12.9 \pm 4.4$\,min.}
    \label{fig:example_flare}
\end{figure}

\begin{figure}
    \centering
    \includegraphics[width=\columnwidth]{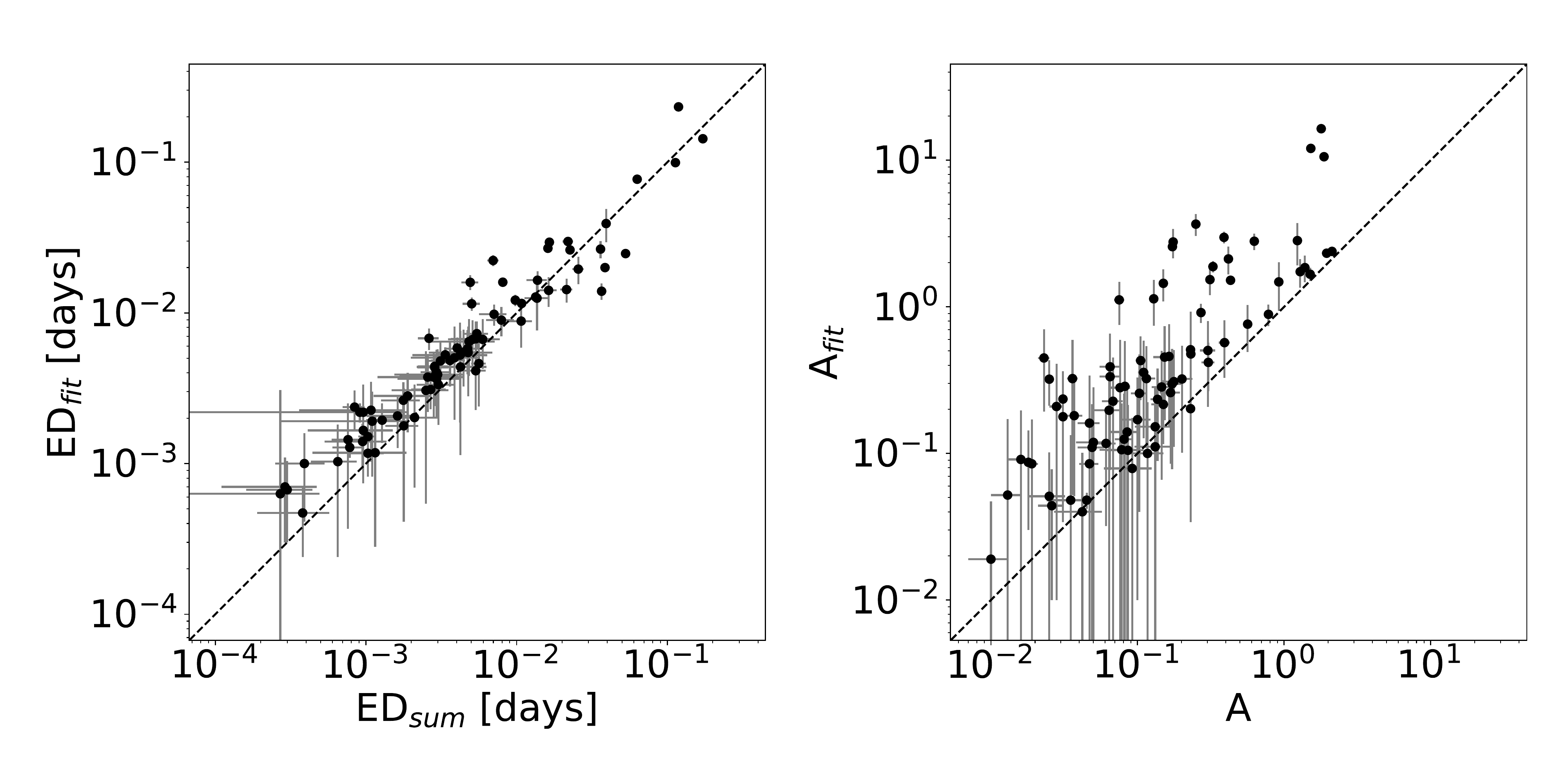}
    \caption{\textit{Left:} Comparison of the summed and fitted EDs. While they generally give the same result, they can differ by a factor of 2--3. \textit{Right:} Comparison of the observed flare amplitude (the highest flux value) and the fitted amplitude. Due to the low time resolution, the observed peak is generally smaller.}
    \label{fig:ED}
\end{figure}

\begin{figure}
    \centering
    \includegraphics[width=\columnwidth]{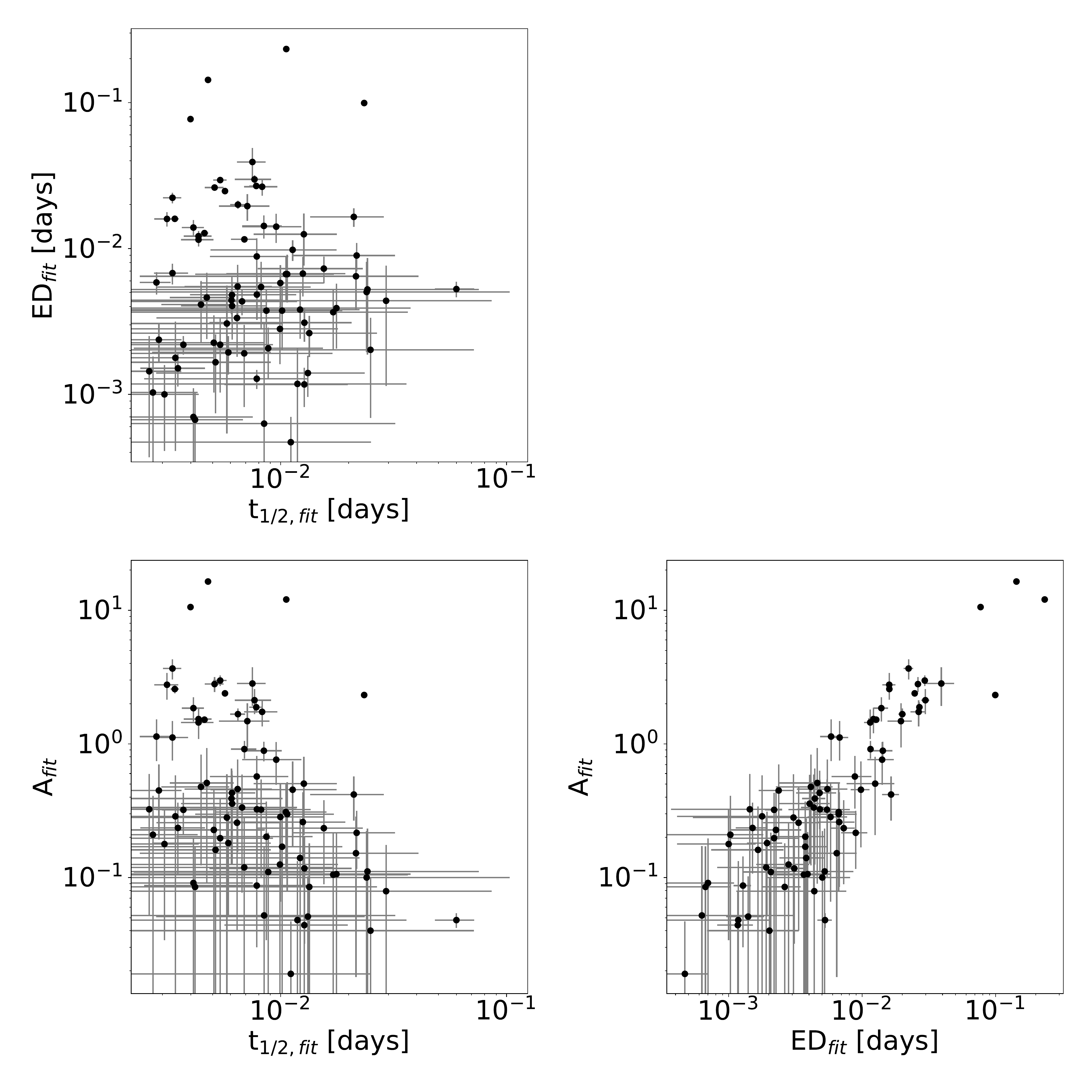}
    \caption{Correlations between the flare parameters.}
    \label{fig:flare_params}
\end{figure}

We run the MCMC chains starting from the maximum a posteriori value for 100~000 steps, discarding the first 20~000 steps as burn-in and only leaving every second step as thinning. A successful fit can be seen in Fig.~\ref{fig:example_flare} (following the example of Fig.~\ref{fig:pca_reconstruction}), while for a few smaller flares the MCMC chain did not converge (13 of the 94 flare events, see Table~\ref{table:flares}). The fits generally overestimated the flare amplitudes (see Fig.~\ref{fig:ED}), and hence the EDs, but they give a reliable estimate of $t_{1/2}$ even for shorter events. A similar behaviour was observed by \cite{Raetz}, where the amplitudes were systematically higher using \textit{Kepler} short cadence light curve compared to long cadence.
We experimented with fitting only two parameters, after forcing the fitted ED to be equal to the previously calculated one, but essentially the same behaviour was observed.
According to Fig.~\ref{fig:flare_params}, the flare EDs mainly scale with the amplitude, and there is no dependence on $t_{1/2}$.

The ED of the 94 flare events were converted to energy by multiplying by the quiescent stellar luminosity. Since the sample consists of stars similar to TRAPPIST--1, we simply used its luminosity for all targets. It was calculated by convolving the \textit{TESS} response function\footnote{\url{https://heasarc.gsfc.nasa.gov/docs/tess/data/tess-response-function-v2.0.csv}} with a BT-NextGen model spectrum ($T_\mathrm{eff} = 2600$\,K, $\log{g} = 5.0$, [Fe/H] = 0), scaling with the $L_\mathrm{bol} = 0.00061 \pm 0.00002 L_\sun$ bolometric luminosity of TRAPPIST--1 from \cite{t1_reanalysis}. The estimated luminosity of TRAPPIST--1 in the \textit{TESS} bandpass is thus  $L_{\mathrm{TESS}} = (2.1 \pm 0.3) \times 10^{29}$\,erg\,s$^{-1}$, with nominal uncertainty calculated using models with $\pm 100$\,K temperature. Using a simple black body spectrum with the same temperature would give $2.9 \times 10^{29}$\,erg\,s$^{-1}$. Integrating the same BT-NextGen model spectrum gives $L_{\mathrm{Kepler}} = (5.6 \pm 1.5) \times 10^{28}$\,erg\,s$^{-1}$ for the luminosity in the \textit{Kepler} bandpass. Compared to the nominal errors cited here, the dominating source of uncertainty is the different luminosities within the sample itself, since $M_\mathrm{G}$ varies by $\pm 0.\!^{\rm m}5$.

As a validation for the temperature and luminosity used above, \cite{t1_reanalysis} lists 4 field dwarfs similar to TRAPPIST--1, with 2603\,K and 0.00058\,$L_\sun$ mean temperature and luminosity in their Table~3, plotted in their Fig.~2a. These 4 stars are also present in our sample, and LHS 132 (Gaia DR2 4989399774745144448) and LHS 3003 (Gaia DR2 6224387727748521344) have already been observed by \textit{TESS}.

\subsubsection{Feasibility check with downgraded \textit{Kepler} K2 data}
\label{sec:k2_downgrade}

\begin{figure}
    \centering
    \includegraphics[width=\columnwidth]{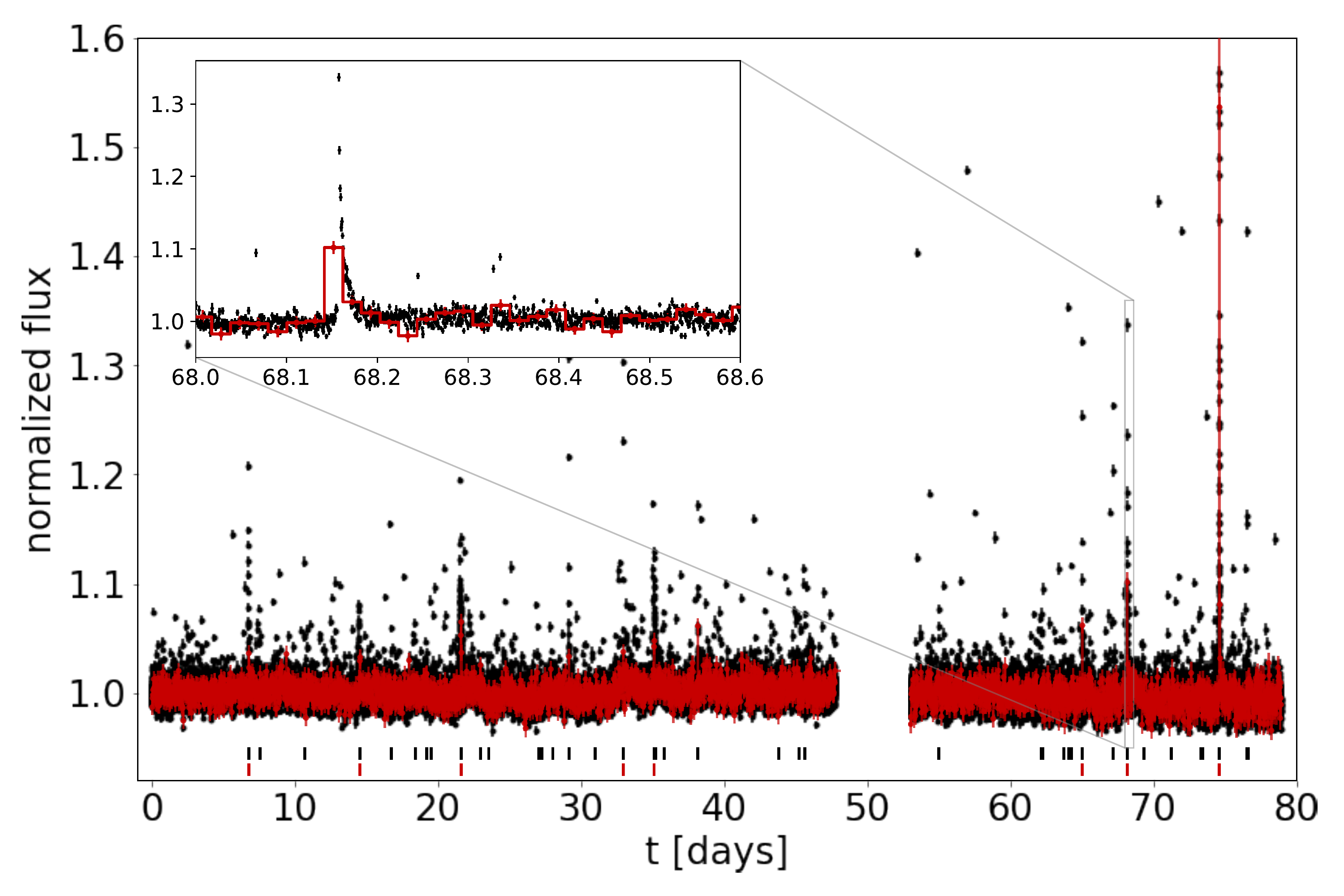}
    \caption{The \textit{Kepler} K2 light curve of TRAPPIST--1 from \cite{t1-flares} (black) rebinned to 30-min time resolution, and with \textit{TESS}-like scatter (red). Black and red ticks below the light curve indicate the position of flares found in the corresponding dataset.}
    \label{fig:kepler_tess_rebin}
\end{figure}

To demonstrate the viability of this project, we converted real photometric observations of TRAPPIST--1 to match the quality of a \textit{TESS} FFI dataset. We took the \textit{Kepler} K2 short cadence light curve of \cite{t1-flares}, converted it from magnitude to flux, scaled up the noise level by a factor of 10 (ratio of the \textit{Kepler} and \textit{TESS} apertures), and rebinned it from 1 to 30-min cadence. We then inspected the resulting light curve (Fig.~\ref{fig:kepler_tess_rebin}), and identified flare events passing the same criteria as the TRAPPIST--1 analogues. While the original light curve showed 42 distinguishable flares, from the downgraded light curve we could only identify 8 of them, only those with the highest energies. However, the energy estimates agreed within the nominal uncertainties. We note that some of the larger flares remained as a single outlier point, but in real observations we could not safely label such events as flares, so we omitted them here as well. It appears that flares with energies above $\sim 10^{31}$\,erg can be observed even with the 10\,cm aperture of \textit{TESS}, on a $T \sim 14^{\mathrm{m}}$ star.

\subsubsection{Completeness}
\label{sec:completeness}

Less energetic flares are harder to distinguish from noise, so they are easily missed. To de-bias the flare frequency distribution (FFD) from this selection effect, we calculate the flare recovery rate (hereafter called completeness) for each light curve, for given flare energies. To do this, we inject artificial flare signals into the light curves, and try to recover them.

Each of the light curves were converted to flux, and clipped with a $3 \sigma$ threshold. The clipped values were then linearly interpolated to fill the gaps induced by real flare events. Next, we added artificial flares with different EDs. To generate a flare with the given ED, $t_\mathrm{peak}$ was drawn from the observed time interval, $t_{1/2}$ was drawn uniformly from 0.001$^\mathrm{d}$ to 0.05$^\mathrm{d}$, and the amplitude of the flare was scaled to match the required ED. Next, the local scatter was calculated in a 4$^\mathrm{d}$ long interval, and the light curve was smoothed with a 0.2$^\mathrm{d}$ wide LOWESS kernel. A flare is considered to be detected if a point is twice the scatter above the smoothed light curve, with the following point still exceeding once the scatter. This criterion was chosen to match visual inspection. For each light curve, 100 ED bins were defined logarithmically from 10 to $10^4$\,s, and 100 random flares were generated for each ED. The completeness is defined as the fraction of recovered flares.

\begin{figure}
    \centering
    \includegraphics[width=\columnwidth]{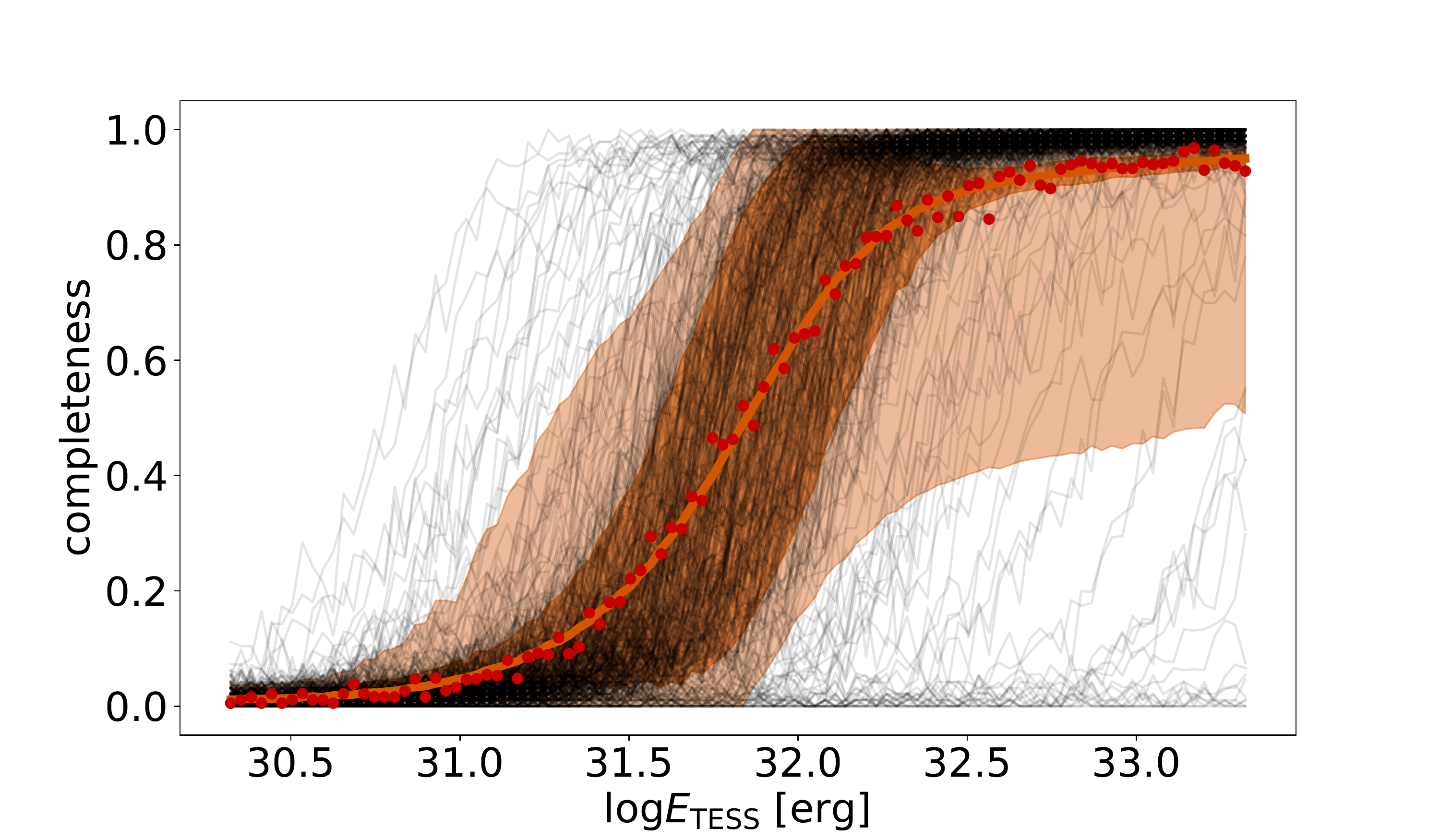}
    \caption{Flare recovery rate calculated for each light curve. The orange line shows the average curve for all stars, red points show the completeness curve calculated for a concatenated light curve. The shaded regions show the 1 and $2 \sigma$ confidence intervals.}
    \label{fig:completeness}
\end{figure}

Taking the multiplicative inverse of the completeness value for each detected flare energy, the FFD calculated from a single light curve can be corrected. However, most of the flares in our sample come from different stars with different photometric noise properties, i.e. with different completeness curves. In order to correct the composite FFD of the whole sample, a single representative correction curve is needed. For this purpose we simply average the completeness curves calculated from each individual light curve. As it can be seen in Fig.~\ref{fig:completeness}, this approach is equivalent to a completeness curve calculated from all the light curves concatenated into a single dataset, spanning $\sim 23$ years. The uncertainty is calculated from the 84th and 16th percentiles of all the single-sector completeness curves. We note that for lower energies even a small difference in the recovery rate can result in a large difference in the composite FFD, thus the lower energy part should be treated with caution. Normally, when a similar approach is used, the difference between the original and corrected FFD is only a few percent. We also note that this method is not strictly correct, as we identified flares manually without a rigorous detection algorithm, since the high noise level of the light curves did not allow us to use one.

The same method was also used to calculate the flare recovery rate for the K2 short cadence light curve of TRAPPIST--1 from \cite{t1-flares}.

The reliability of the ED estimation method was also tested in a similar manner, by injecting artificial flares into the light curves. Due to the high photometric noise, the measured differences in ED can exceed the nominal uncertainties. For flares above $E_\mathrm{TESS} = 10^{32}$\,erg, the measured EDs were erroneous by a factor of 10 in $\sim 5$\% of the injected flares. However, the differences are not expected to be this severe for most of the detected events, since we have used all light curves -- even the noisiest ones -- for the injection tests.

\subsection{H$\alpha$ emission}
\label{sec:halpha}

To look for emission features, we checked the \textit{Rizzo Spectral Library}. 95 stars from our sample had available optical spectra, published by \cite{rizzo1}, \cite{rizzo2} and \cite{rizzo3}. The average wavelength range of these spectra is from 5600\,\AA \,to 10000\,\AA, with average $\frac{\lambda}{\Delta \lambda}$ resolution between 2000 and 5000. Most stars showed prominent emission in H$\alpha$. We calculated the H$\alpha$ equivalent width (EW) by integrating the flux between 6553 and 6573\,\AA, after normalising by a linear fit to the local continuum. The uncertainty of the EW was estimated by re-sampling from the flux errors. We note that the flux errorbars were generally larger in the vicinity of H$\alpha$ if the emission was stronger, leading to high (and possibly overestimated) uncertainties even for high EW values. The measured EWs are listed in Table~\ref{table:targets}.

\subsection{Age determination}
\label{sec:age}

To see how the activity signatures evolve, it would be interesting to have an age estimation for as many stars from the sample as possible. While TRAPPIST--1 itself is fairly old ($7.6 \pm 2.2$ Gyr, \citealt{t1-age}), the sample discussed here should include objects with different ages, since these low-mass objects move on the Hertzsprung--Russell diagram slowly during their evolution, while the selection criteria were related to colour and luminosity alone. Unfortunately, all the other parameters of these stars evolve slowly with age, making the inference troublesome. In this section we explore the available methods to estimate stellar ages in this low-mass regime.

While the ages of individual objects are hard to estimate, it is not the case for clusters and moving groups. With available astrometric information the probability that these stars are members of known moving groups can be estimated. To this end, we utilized BANYAN $\Sigma$ \citep{banyan}. It can predict membership probabilities for 27 young associations within 150\,pc, using position and velocity information. No stars in the sample had measured radial velocity in \textit{Gaia} DR2, but 72 of them had literature measurements in the Simbad database. For these stars all 6-dimensional position and velocity information was used, while the radial velocity was omitted for the remaining objects. 14 stars had membership probability over 95\%. \textit{Gaia} DR2 65638443294980224, 3200303384927512960, 4900323420040865792 and 89186168428165632 had high membership probability and confirmation from the literature (\cite{2013A&A...559A..43G}, \cite{baron2019}, \cite{gagne2015} and \cite{SpeX}, respectively). \cite{burgasser2015} have found that \textit{Gaia} DR2 6224387727748521344 does not show signs of young age, so it was rejected as an AB Doradus member. Although the remaining stars have no confirmation of young age from the literature nor measured radial velocity, they were kept as candidate moving group members.

For those 72 stars that had radial velocity measurements the full 6-dimensional position and velocity information is available. This lends itself to kinematic age estimation, a method that uses the motion within the galactic disk as a proxy for age. Young objects tend to have simple orbits with smaller peculiar velocities, while older stars are scattered to more eccentric and inclined orbits. We used the method of \cite{kinematic_age} to calculate the age probability density function (PDF) from the UVW galactic velocity components. The method is calibrated using isochronal ages of $\sim$14~000 stars from the Geneva--Copenhagen Survey. While most of the other methods (like isochrone fitting) is not applicable to low-mass stars, there is no mass dependency in the kinematic method. As a downside, the precision is low (the $1 \sigma$ errorbars are typically 2--3\,Gyr).

\begin{figure}
    \centering
    \includegraphics[width=\columnwidth]{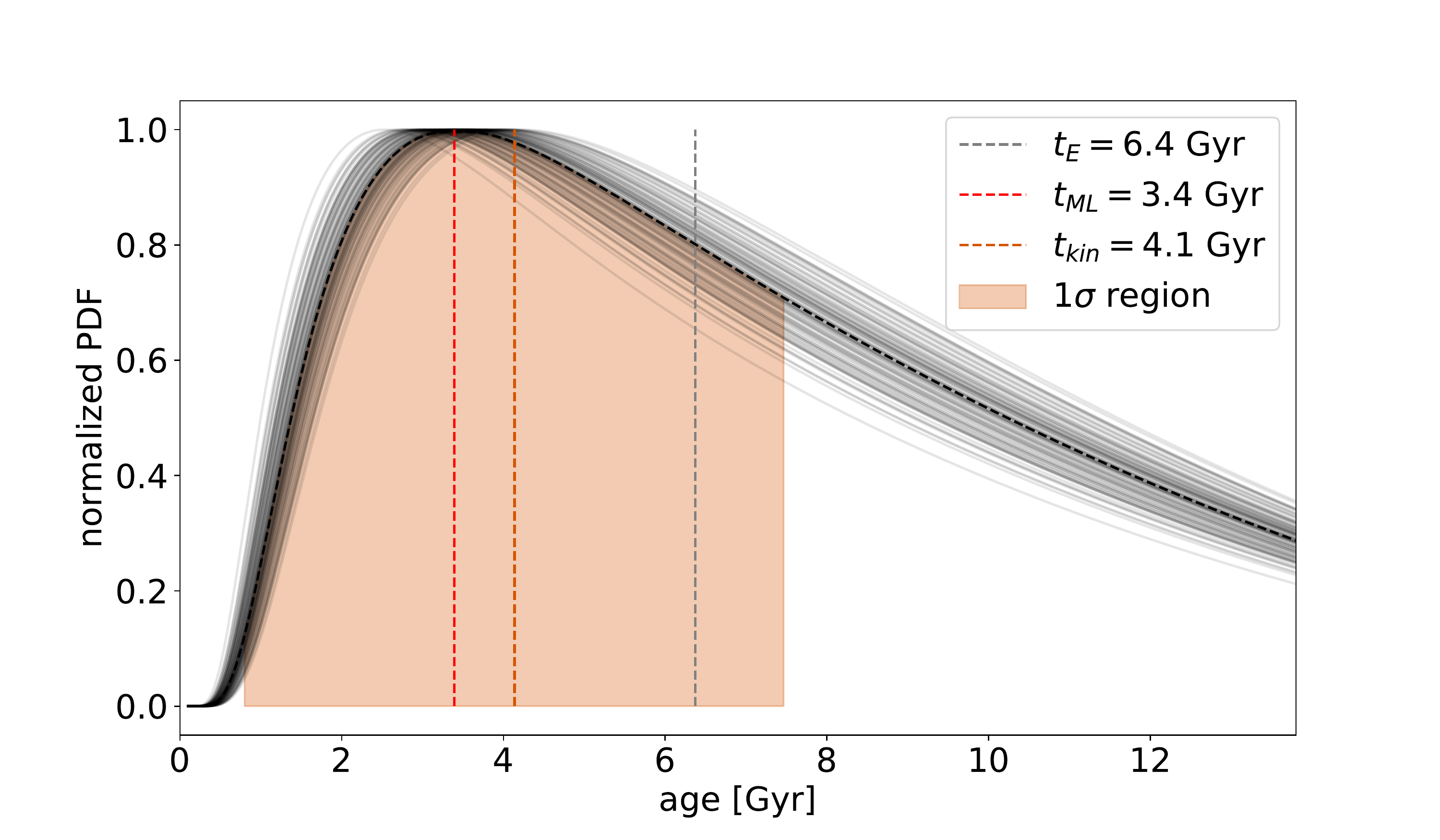}
    \caption{The kinematic age PDF of \textit{Gaia} DR2 3421840993510952192, with $(U,V,W) = (33.0 \pm 4.0, -20.5 \pm 0.2, -22.2 \pm 0.5)$\,km s$^{-1}$. Thin black lines show 100 realizations of the PDF taking into account measurement errors.}
    \label{fig:kinematic_age_PDF}
\end{figure}

First we converted the \textit{Gaia} DR2 position, proper motion, parallax and literature radial velocity data to galactic velocity components. The measurement uncertainties were propagated into the UVW components by re-sampling the input values. We then used the "UVW method" described in Sect.~3.1 of \cite{kinematic_age}, by calculating the components of the velocity ellipsoid, and building the age PDF as a product of 3 Gaussians. To extract a point estimate from the PDF, we calculated the most likely age ($t_\mathrm{ML}$) as the mode of the distribution, and the expected age ($t_\mathrm{E}$) as
\begin{equation}
    t_\mathrm{E} = \int_{0}^{13.8\,Gyr}{t p(t|U,V,W) dt}
\end{equation}
where $p(t|U,V,W)$ is the age PDF calculated numerically between 0 and 13.8\,Gyr. Following the suggestions of \cite{kinematic_age}, we used the following kinematic age as the estimated value:
\begin{equation}
    t_\mathrm{kin} = \frac{3 t_\mathrm{ML} + t_\mathrm{E}}{4}
\end{equation}
with the uncertainty of
\begin{equation}\label{eq:delta_t}
    \delta_t = \frac{1}{4} \left[ (t_{84} - t_{16}) + \frac{t_{97.5} - t_{2.5}}{2} \right]
\end{equation}
where the subscripts of $t$ denote percentiles. In the case of Gaussian distribution $\delta_t$ corresponds to $1 \sigma$. To take the measurement uncertainty of the UVW components into account, 100 realizations of the PDF were calculated using re-sampled UVW values, and the point estimates were calculated from the median PDF. We also chose to add the scatter in the 100 individual $t_\mathrm{kin}$ values in quadrature to $\delta_t$, which generally increased $\delta_t$ by $\sim 0.1$\,Gyr. To illustrate the method, an example PDF can be seen in Fig.~\ref{fig:kinematic_age_PDF}.

An other technique to determine ages of UCDs is to look for wide binary pairs, where the age of the other component can be measured (assuming coevality). For all stars in the sample, we queried \textit{Gaia} DR2 looking for stars with similar astrometric parameters. For this we utilized the same criteria as proposed by \cite{gaia_proper_motion_match}, except that our search was limited to a projected separation of 0.2\,pc. This resulted in 19 comoving pair candidates, with 0.05\,pc maximum separation, summarised in Table~\ref{table:comoving_pairs}. Radial velocity measurement was available for 9 stars, making kinematic age estimation possible.

Out of these 19 stars, \textit{Gaia} DR2 1412377317863375488 has measured rotational period, but no kinematic age estimate. Its comoving pair is HD~234344, a bright K3 dwarf with measured radial velocity from \textit{Gaia} DR2. The kinematic method gives $3.7 \pm 3.3$\,Gyr, while isochrone fitting combined with gyrochronology with \textsc{stardate} \citep{gyrochronology} gives $0.1_{-0.1}^{+2.4}$\,Gyr (using the peak of the age posterior). We used $3.1853^\mathrm{d} \pm 0.004^\mathrm{d}$ rotational period measured from 3 sectors of \textit{TESS} FFI data.

As an independent means of age estimation, we inspected all the available optical spectra from the \textit{Rizzo Spectral Library} looking for lithium absorption around 6708\,\AA, and found none. The lithium depletion occurs around 100--200\,Myr for stars in the mass range of 0.06--0.08\,$M_\sun$ \citep{burke_lithium}, thus giving a lower limit for the age of these 95 stars. This rules out \textit{Gaia} DR2 2755265775727402112, as BANYAN $\Sigma$ gives 97\% Columba membership probability, while the age of the association is only 42\,Myr \citep{columba_association}, thus the star should still show lithium absorption.

\begin{figure}
    \centering
    \includegraphics[width=\columnwidth]{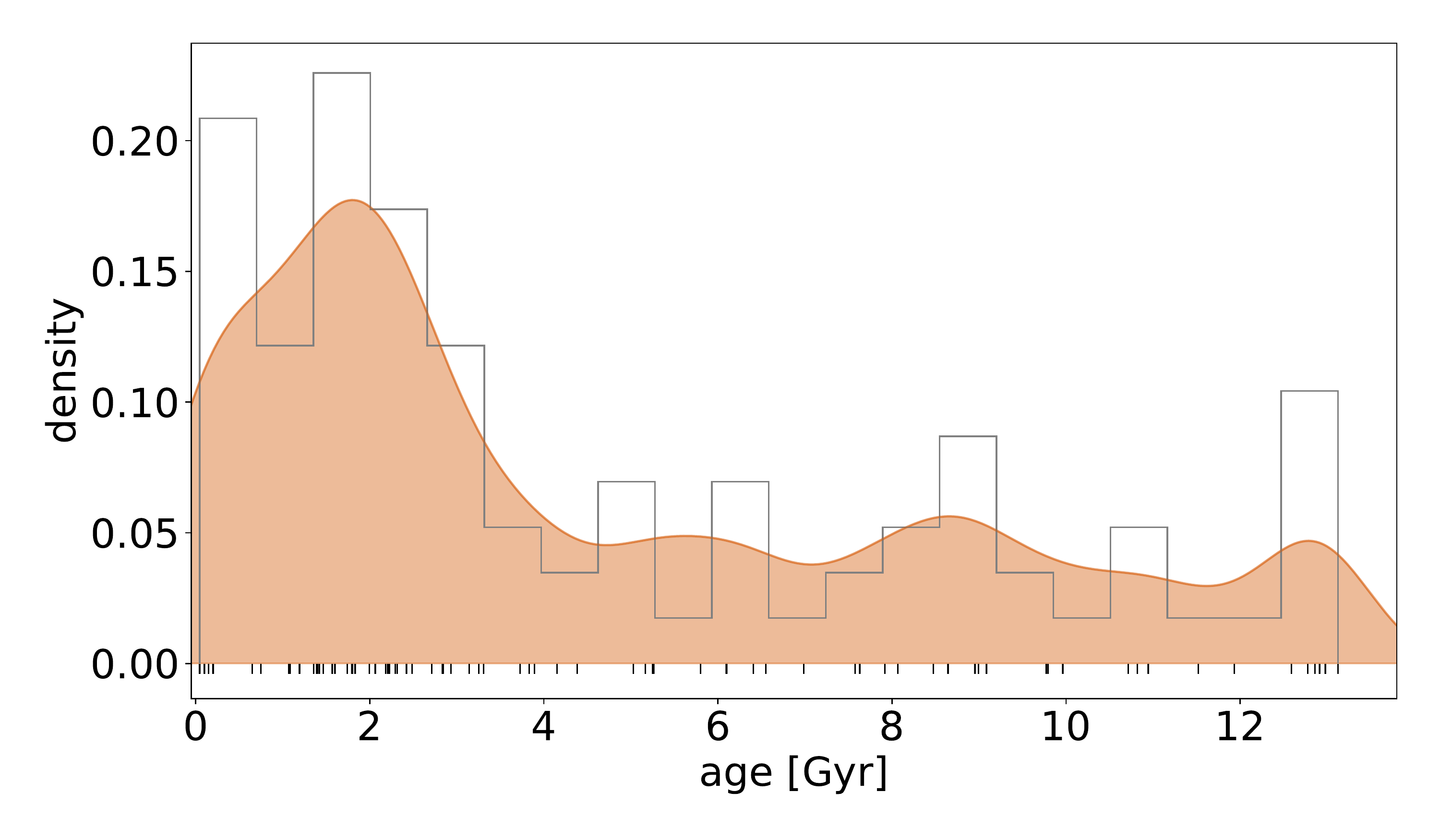}
    \caption{The distribution of the ages of the stars in the sample. The kernel density estimation is calculated with Gaussian kernel with 0.6\,Gyr bandwidth. Ticks below the curve show the individual data points.}
    \label{fig:age_kde}
\end{figure}

\begin{figure*}
    \centering
    \includegraphics[width=\textwidth]{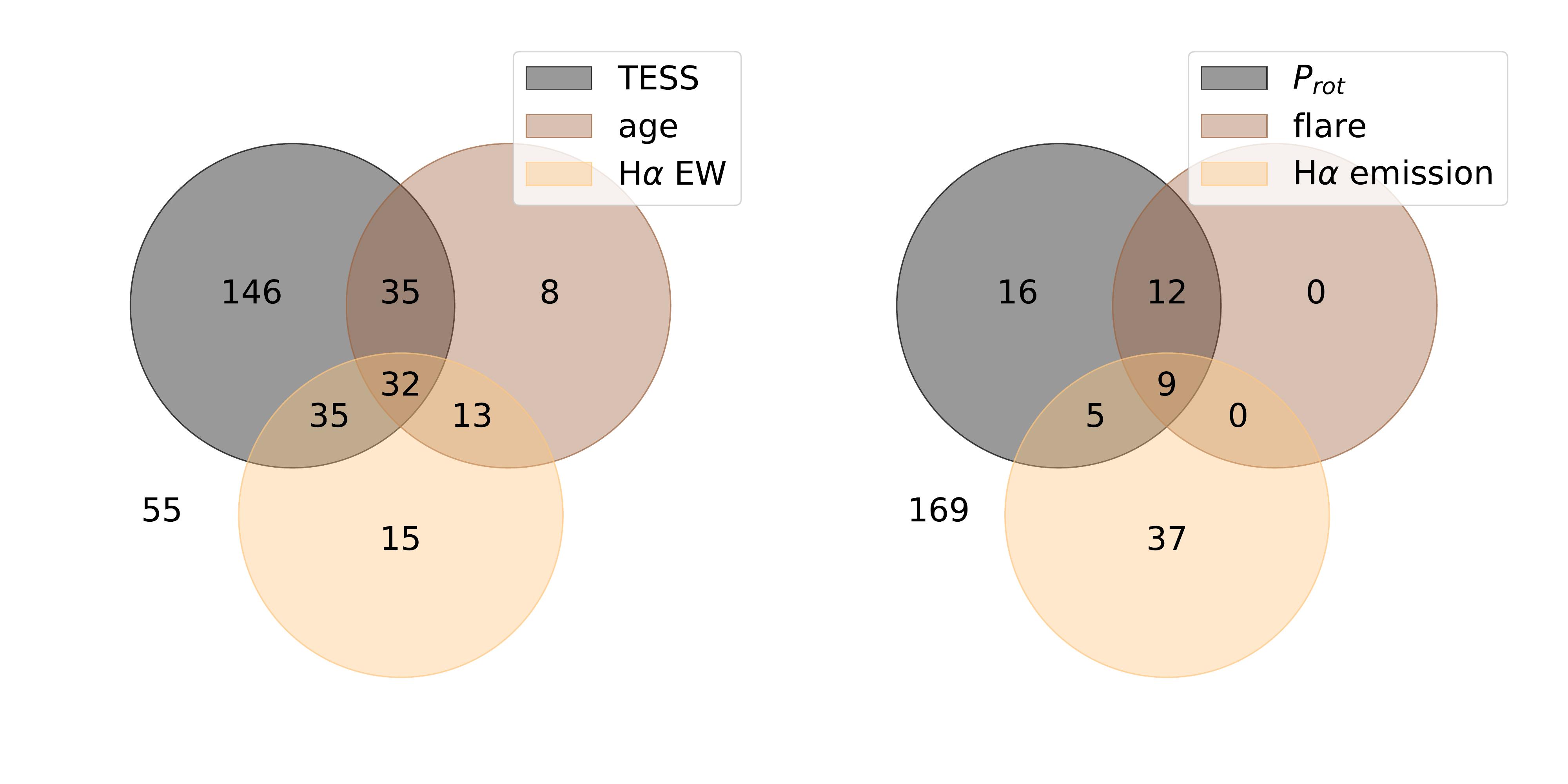}
    \caption{Venn diagrams with the main properties of our sample. \textit{Left:} The number of stars with available \textit{TESS} light curve, age estimate and calculated H$\alpha$ EW from the whole sample of 339 stars. \textit{Right:} The number of stars with detected period, flaring and H$\alpha$ emission (where the EW is at least $1 \sigma$ above zero), only for the 248 stars with available \textit{TESS} observation.}
    \label{fig:venn_diagram}
\end{figure*}

To summarise, we have age estimates from the following sources (prioritizing them in the following order, removing intersections): 12 from moving group membership, 5 from comoving pairs and 71 from kinematics. These are tabulated in Table~\ref{table:targets}, and their distribution is plotted in Fig.~\ref{fig:age_kde}. The number of stars in our sample with different estimated parameters are shown in Fig.~\ref{fig:venn_diagram}.

\section{Results and discussion}
\label{sec:discussion}

\subsection{Notes on individual stars}

\begin{figure}
    \centering
    \includegraphics[width=\columnwidth]{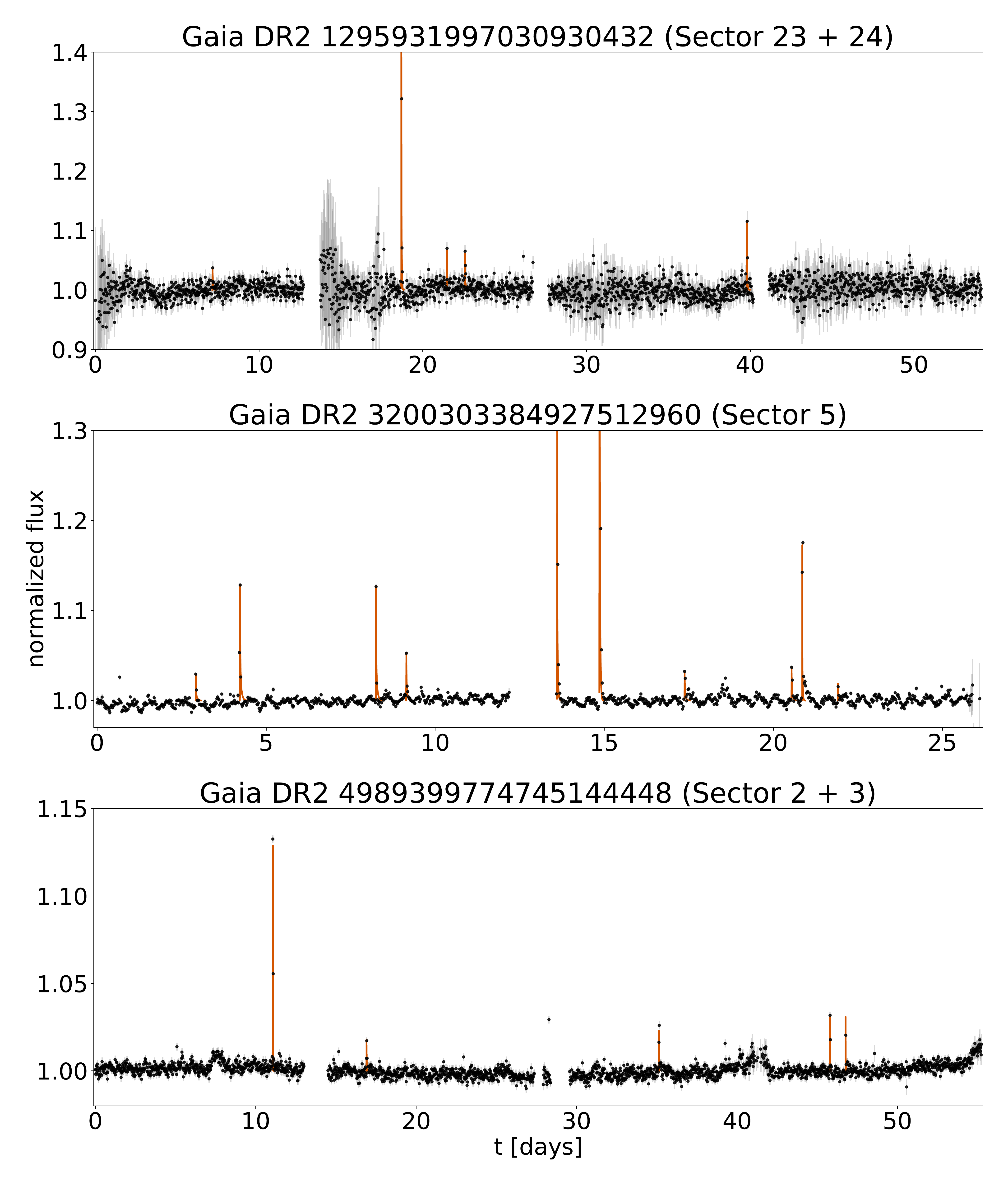}
    \caption{The three most flaring targets from the sample, with $G_\mathrm{RP} = 15.\!^{\rm m}8$, $13.\!^{\rm m}4$ and $13.\!^{\rm m}8$, respectively. The detected flares are marked with orange.}
    \label{fig:individual_stars}
\end{figure}

 The most flaring stars of the sample are \textit{Gaia} DR2 1295931997030930432 (5 flares), 3200303384927512960 (10 flares) and 4989399774745144448 (5 flares). Figure~\ref{fig:individual_stars} shows the available \textit{TESS} light curves. Taking into account their different apparent magnitudes, these stars show a flaring rate comparable to TRAPPIST--1, as the downgraded K2 light curve from Sect.~\ref{sec:k2_downgrade} yielded 8 flare events in $70^\mathrm{d}$ of observation. While there is no trace in the literature of \textit{Gaia} DR2 1295931997030930432, we know a bit more about the other two stars. \textit{Gaia} DR2 3200303384927512960 (= 2MASS J04402325-0530082) appears in the sample of surveying nearby M-dwarfs searching for brown dwarfs in \cite{2020A&A...634A.128N} where a SED fit is presented with $T_\mathrm{eff} = 2600$\,K and $\log{g} = 4.5$. \textit{Gaia} DR2 4989399774745144448 (= 2MASS J01025100-3737438) is listed as flaring star by \cite{2019ApJ...870...10M} using observations from the MEarth photometric survey. Finally, we note that the well-known and observed flare star \textit{Gaia} DR2 4339417394313320192 = vB8 (2MASS J16553529-0823401), which has a measured magnetic field ($2.8 \pm 0.4$\,kG, \citealt{2019A&A...626A..86S}), is also included in our sample, but was not observed by TESS yet.

\subsection{Periods}

Photometric observations of TRAPPIST--1 yielded different values for its rotational period: $0.819^\mathrm{d}$ from Spitzer infrared data \citep{t1-rachael}, $1.60^\mathrm{d}$ from TRAPPIST-South \citep{t1_discovery}, $3.30^\mathrm{d}$ from K2 \citep{t1_period_luger,t1-flares}. The reason for the discrepancy could be the changing pattern of spots on the surface that could imitate e.g. half or one-fourth of the real period. Such effect of quickly evolving spot pattern was not observed in our sample, mainly due to the relatively short observing time. Out of the 42 stars with periodic signal, \textit{Gaia} DR2 1656001233124961152 was observed for the longest time (11 sectors). To find potential evidence of changing spot pattern, we compared the Lomb-Scargle periodograms of each sector, and employed Short Term Fourier Transform (see e.g., \citealt{stft}). While the main period was only detected in 7 out of the 11 sectors, it is unclear whether it stems from some change in spot configuration, or just data systematics and noise. The half of this period was not detected in any of the sectors.

The histogram of the detected periods of TRAPPIST--1 analogues is plotted in Fig.~\ref{fig:kde}. The distribution seems to be unimodal and follow a log-normal distribution. We performed a Shapiro--Wilk test \citep{shapiro_test} on the logarithm of the periods, and it could not reject the null hypothesis of normal distribution ($p=0.37$). As a test for unimodality, we fitted the $\log{P_\mathrm{rot}}$ dataset with a mixture of $N$ Gaussians. The $N=1$ model was clearly preferred by the Bayesian Information Criterion (with $\mu = 0.6^\mathrm{d}$ mean and $\sigma = 0.6^\mathrm{d}$ scatter). To put this into context, we compare it with other distributions from the literature. We try to separate each distribution using Gaussian Mixture models with the number of components determined from the Bayesian Information Criterion. We note that we use the logarithm of the periods, but transform the values back to make comparison easier.

The period distribution of early-to-mid M-dwarfs in the \textit{Kepler} field was drawn first by \citet{2013MNRAS.432.1203M} based on 10 months of data. While they found a distinct group of shorter period stars making up $\sim 8$\% of the full sample, the most striking feature is the emergence of bimodality for longer periods, possibly arising from two distinct waves of star formation. The best Gaussian Mixture fit is achieved with the following 3 components: $(\mu_1, \sigma_1) = (3^\mathrm{d}, 3^\mathrm{d})$ for the fast rotators, and $(\mu_2, \sigma_2) = (18^\mathrm{d}, 4^\mathrm{d})$, $(\mu_3, \sigma_3) = (34^\mathrm{d}, 7^\mathrm{d})$ for the longer periods. Later, \citet{m_dwarf_periods} using all 16 quarters of \textit{Kepler} data detailed the period distribution of 297 M-dwarfs in the short period range. A two component fit to their data yields $(\mu_1, \sigma_1) = (0.4^\mathrm{d}, 0.2^\mathrm{d})$, $(\mu_2, \sigma_2) = (1.2^\mathrm{d}, 0.4^\mathrm{d})$. Using ground-based photometry from the MEarth Project, \cite{mearth_periods1, mearth_periods2} compiled the rotational period of $\sim600$ mid-to-late M-dwarfs, also finding bimodality in the distribution. Fitting a Gaussian Mixture model to the logarithm of Grade A and B rotational periods from \cite{mearth_periods1, mearth_periods2} gives the following parameters: $(\mu_1, \sigma_1) = (0.6^\mathrm{d}, 0.5^\mathrm{d})$ for the fast rotators, $(\mu_2, \sigma_2) = (93^\mathrm{d}, 25^\mathrm{d})$ for the slow rotators, and a wide (possibly background) component in between: $(\mu_3, \sigma_3) = (6^\mathrm{d}, 9^\mathrm{d})$.

These support that the cutoff in Fig.~\ref{fig:kde} is not entirely due to the 5$^\mathrm{d}$ upper limit in our period detection method, but there really is a dearth of objects with periods above a few days. Also the distribution seems to be consistent with the MEarth dataset, even with the truncation. It is likely that many stars from our sample belong to the slow rotating group ($P_\mathrm{rot} \sim 100^\mathrm{d}$), making one or two sectors of \textit{TESS} data insufficient to reliably measure their periods.

\begin{figure}
    \centering
    \includegraphics[width=\columnwidth]{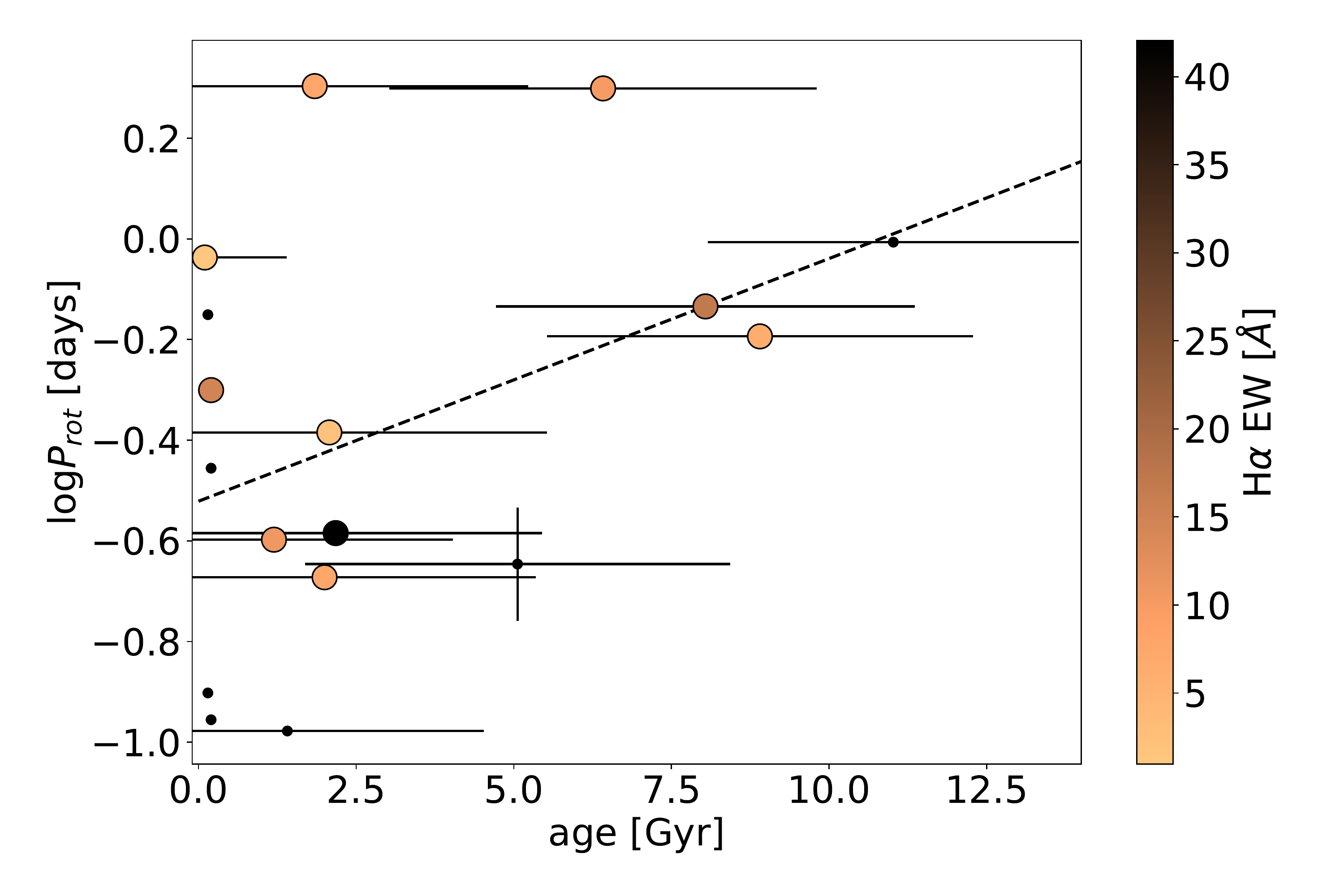}
    \caption{Age and rotational periods for the sample. Dashes line shows a linear fit, small black points show stars without H$\alpha$ EW measurement.}
    \label{fig:age_prot}
\end{figure}

Figure~\ref{fig:age_prot} shows really weak correlation between $\log{P_\mathrm{rot}}$ and age, if any (0.43 Pearson correlation coefficient, with $p=0.09$). There are several ways how $P_\mathrm{rot}$ can evolve (see \citealt{rotation_theory} for an overview), one of which is the initial spin up due to contraction. However, the stars in our sample are likely older than the typical spin up timescale, as illustrated in Fig.~\ref{fig:period_evolution}. After reaching the main sequence, UCDs evolve slowly, losing angular momentum with stellar wind. While this mechanism is effective for solar-type stars, UCDs spin down slowly, retaining their fast rotation for Gyr timescales (see e.g., \citealt{ang_mom_evolution, mearth_periods1}). \cite{irwin_ang_mom_evolution} studied a sample of 41 fully convective field M-dwarfs with $P_\mathrm{rot}$ measured from MEarth data. In their Fig.~13, they show age vs $P_\mathrm{rot}$ for their sample and also for several open clusters. Viewing only stars older than $\sim 10$\,Myr and with $P_\mathrm{rot} > 10^\mathrm{d}$ gives a similarly large scatter as Fig.~\ref{fig:age_prot} here.

\begin{figure}
    \centering
    \includegraphics[width=\columnwidth]{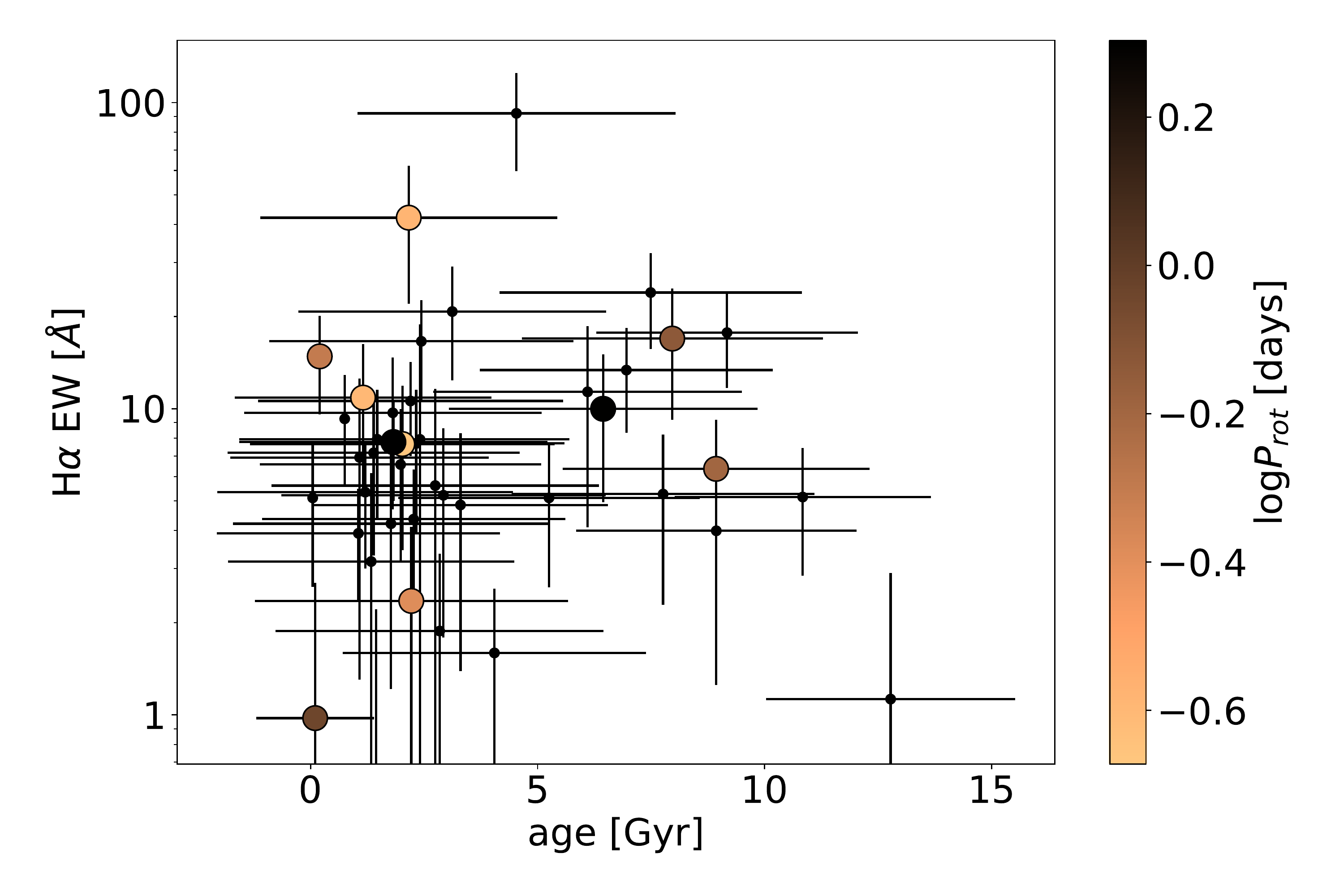}
    \caption{Age and H$\alpha$ emission. There appears to be no clear trend with the rotational period.}
    \label{fig:age_halpha}
\end{figure}

\begin{figure}
    \centering
    \includegraphics[width=\columnwidth]{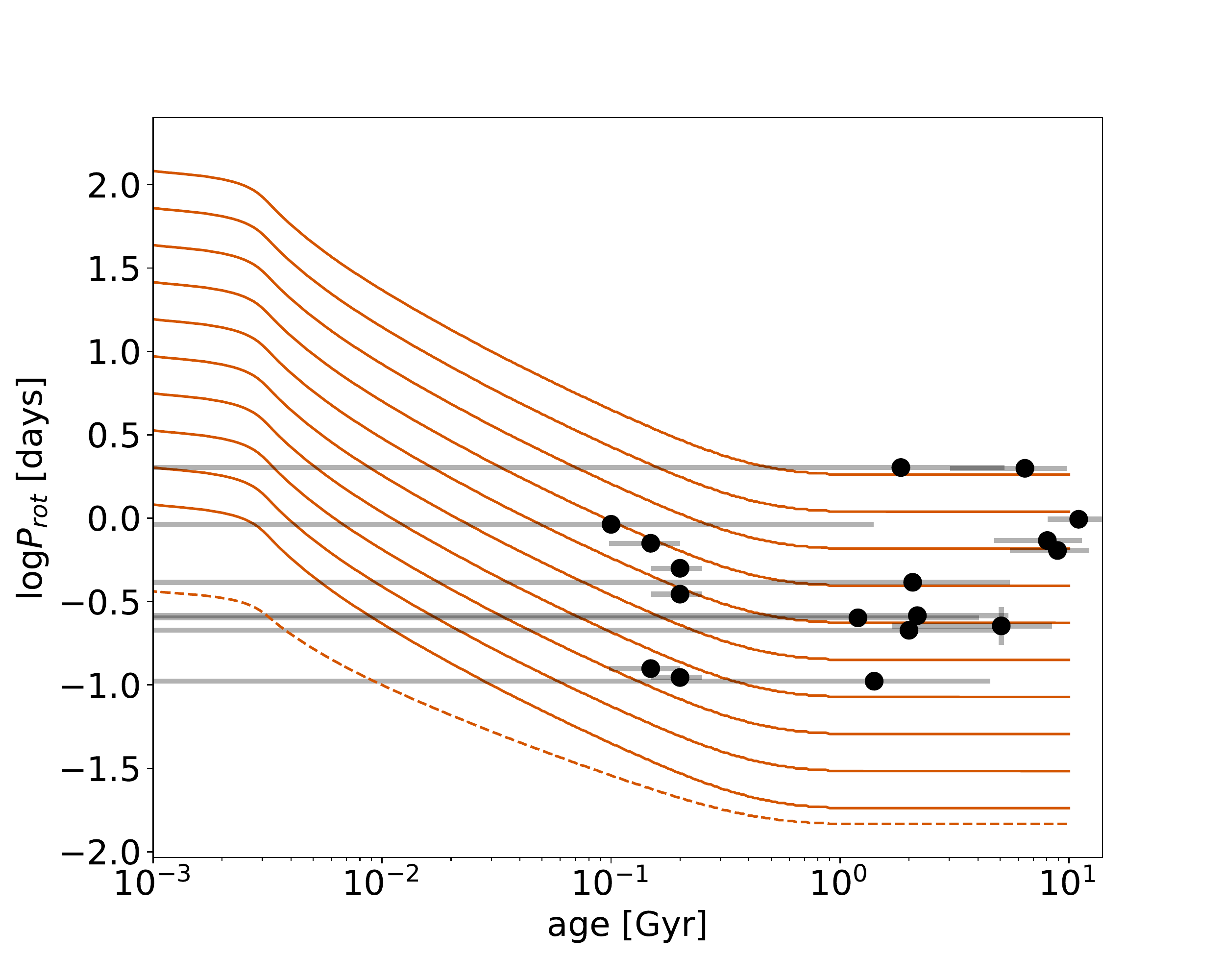}
    \caption{Rotational period vs age for the TRAPPIST--1 analogues, compared with $M = 0.09$\,$M_\odot$ evolutionary tracks from \cite{baraffe}, assuming angular momentum conservation with different initial $P_\mathrm{rot}$ values. The dashed line shows the breakup period $P_\mathrm{crit} [\mathrm{days}] = 0.116 (R / R_\odot)^{3/2} (M / M_\odot)^{-1/2}$ from \cite{breakup_period}.}
    \label{fig:period_evolution}
\end{figure}

According to Fig.~\ref{fig:age_halpha}, the H$\alpha$ emission does not seem to correlate with age or $P_\mathrm{rot}$. A possible explanation could be the saturated activity in these late type stars. \cite{t1-rachael} showed that the Rossby number $R_0 = P_\mathrm{rot} / \tau_c$ is below 0.1 for TRAPPIST--1, and in fact for all stars with measured $P_\mathrm{rot}$ in our sample, taking $\tau_c = 70^\mathrm{d}$ for the convective turnover time (following \citealt{rossby_number_70}). This means that the rotation rate no longer influences the activity of the 42 fast rotators found here. It can be seen in Fig.~\ref{fig:venn_diagram} that one-fourth of the stars with detected H$\alpha$ emission are fast rotators.

\subsubsection{Flare frequency distribution}

\begin{figure}
    \centering
    \includegraphics[width=0.95\columnwidth]{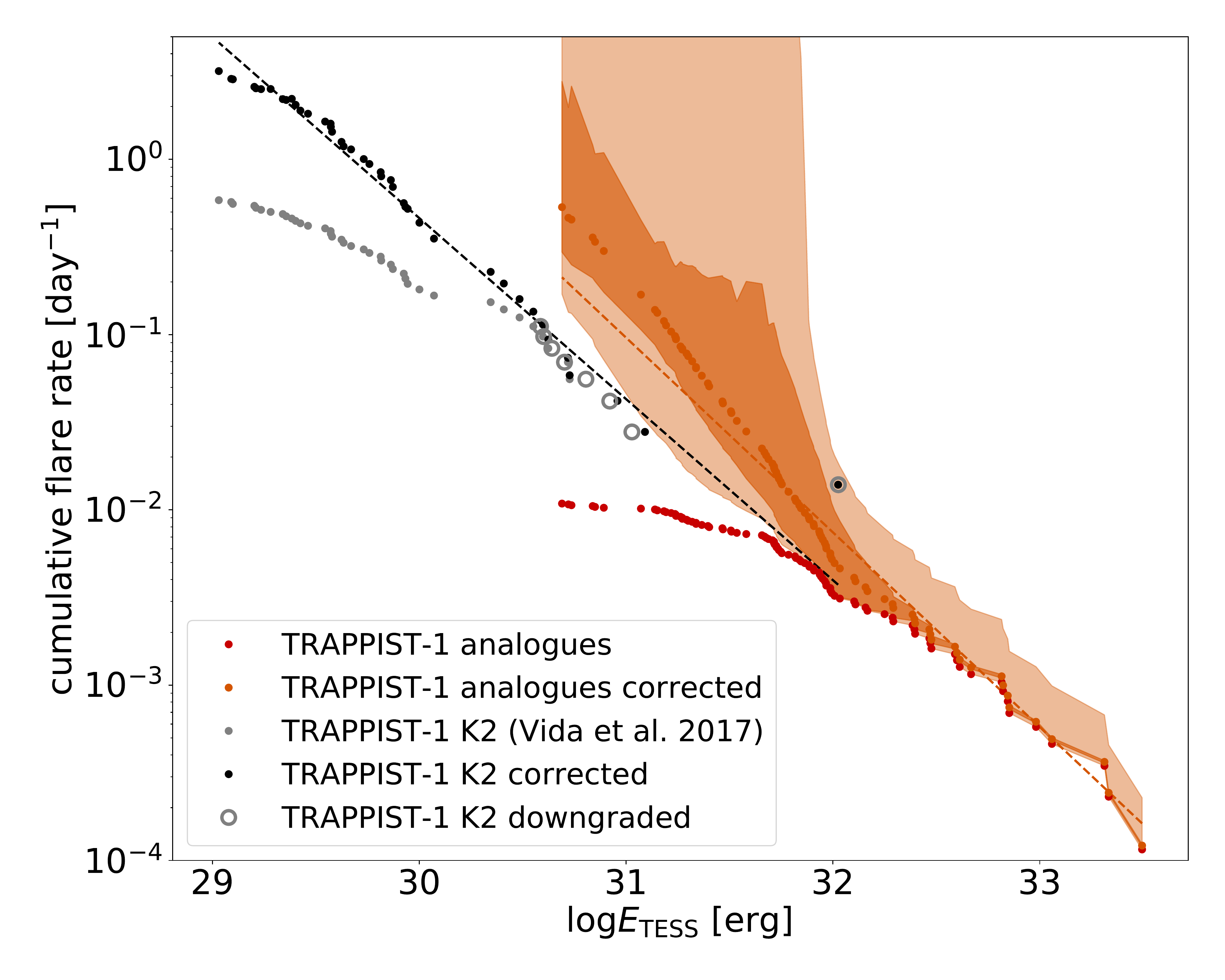}
    \caption{Composite flare frequency distribution for the whole sample. Red points are the original values, orange points are corrected for flare recovery rate (orange line in Fig \ref{fig:completeness}), and shaded regions are the 1 and $2 \sigma$ confidence intervals. The FFD of TRAPPIST--1 with K2 short cadence data from \cite{t1-flares} is also plotted for reference, converted to the \textit{TESS} bandpass, and also its correction for completeness. Grey circles shows the flares recovered from the K2 light curve of TRAPPIST--1 downgraded to imitate a hypothetical \textit{TESS} observation (with added noise, and re-sampled to 30-min cadence). Dashed lines show linear fits.}
    \label{fig:FFD}
\end{figure}

With the estimated flare energies from Sect.~\ref{sec:flares}, the cumulative number of flares above given energies -- the flare frequency distribution (FFD) -- is shown in Fig.~\ref{fig:FFD}. It includes flares from 21 stars, all of which have measured rotational periods (see Fig.~\ref{fig:venn_diagram}). We note that normally the FFD is plotted only for a single star, but here we used the whole sample, since most flaring stars showed only one event. This way, the composite FFD can be treated as an "average" FFD of the whole sample, including stars with different parameters, most notably different ages. However, according to Fig.~\ref{fig:age_prot} and Fig.~\ref{fig:age_halpha} the ages of the stars do not really affect their activity levels, i.e., no clear trend is found between the age and the rotational periods and/or H$\alpha$ emission. The original FFD was corrected by dividing with the averaged completeness curve shown in Fig.~\ref{fig:completeness}. The corrected FFD was fitted with a linear function on log--log scale, yielding
\begin{equation}
    \log(\nu [\mathrm{day}^{-1}]) = (-1.11 \pm 0.02) \log(E_{\mathrm{TESS}} [\mathrm{erg}]) + (33.4 \pm 0.8)
\end{equation}
where $\nu$ is the cumulative flare rate. This results in $\alpha = 2.11 \pm 0.02$ power law index, where
\begin{equation}
    \frac{dN(E_\mathrm{TESS})}{dE_\mathrm{TESS}} \propto E_\mathrm{TESS}^{-\alpha}
\end{equation}
As the uncertainty due to the completeness correction was taken into account in the fit, the lower energy part had only negligible contribution. Fitting only above the break point at $\log{E_\mathrm{TESS}} = 32$ gives $\alpha = 2.11 \pm 0.05$.

\cite{t1-flares} created the FFD of TRAPPIST--1 using short cadence \textit{Kepler} K2 data. To compare the flare energies between \textit{Kepler} and \textit{TESS}, we calculated an approximate conversion factor by assuming $(9000 \pm 500)$\,K black body spectrum for the emitting region \citep{sun_9000k}. We then convolved the spectral response functions of the instruments with the black body spectrum, and integrated over wavelength. Thus, the flare energies from \textit{Kepler} can be converted to flare energies in the \textit{TESS} bandpass as follows:
\begin{equation}\label{E_kep_to_tess}
	E_{\rm TESS} = (0.72 \pm 0.02) E_{\rm Kepler},
\end{equation}
This results in a 0.14 shift on logarithmic energy scale. Similarly, an approximate conversion factor to bolometric flare energy was calculated using a $(9000 \pm 500)$\,K black body:
\begin{equation}
	E_{\rm bol} = (6.0 \pm 0.5) E_{\rm TESS}
\end{equation}
We note that while the \textit{Kepler--TESS} conversion was relatively insensitive to the assumed black body temperature, it is not the case for the bolometric conversion. While $T = 9000$\,K is generally used in the literature for the effective flare temperature (see e.g. \citealt{abiogenesis_conversion}), \cite{evryflare_temperature} demonstrated that superflares can emit at significantly higher temperatures. So the bolometric flare energies should be treated with caution, as they are likely lower limits. Before applying Equation \ref{E_kep_to_tess}, the K2 flare energies were recalculated using $L_{\mathrm{Kepler}} = 5.6 \times 10^{28}$\,erg\,s$^{-1}$ from Sect.~\ref{sec:flares}, since \cite{t1-flares} used a different luminosity. Figure \ref{fig:FFD} also shows the K2 FFD and its downgraded version from Sect.~\ref{sec:k2_downgrade}. It can be seen that the measurements presented here are complementary to TRAPPIST--1, as they probe a higher energy range not observed before. The K2 FFD was also fitted with a line on log--log scale, taking into account the completeness correction from Sect.~\ref{sec:completeness}, and omitting the highest energy point which seems to deviate from the trend, resulting in:
\begin{equation}
    \log(\nu [\mathrm{day}^{-1}]) = (-1.03 \pm 0.02) \log(E_{\mathrm{TESS}} [\mathrm{erg}]) + (30.7 \pm 0.6)
\end{equation}
This yields $\alpha = 2.03 \pm 0.02$ for TRAPPIST--1, a value consistent with the one found for our sample, suggesting that the FFD can be described by a single power law. Not using the correction for recovery rate would give $\alpha = 1.59 \pm 0.02$, a significantly shallower slope.

\cite{hawley_for_ffd} analysed the flare rate of early-to-mid M-dwarfs with \textit{Kepler} short cadence data, and found the steepest power law index $\alpha = 2.32$ on the latest M-dwarf binary in their sample, GJ 1245 AB (M5+M5). Once the light curve of the binary was separated by \cite{lurie_gj1245}, the two independent power law indices changed to $\alpha = 1.99 \pm 0.02$ and $\alpha = 2.03 \pm 0.02$. \cite{k2_ucd} used K2 light curves to study flares on 10 UCDs (including TRAPPIST--1), and found $\alpha$ ranging from 1.34 to 2.04, with 1.66 mean. \cite{k2_ucd_ii} used K2 data to study 3 young UCDs, and found $\alpha = 1.8$. For the mid-to-late M-dwarf sample of \cite{medina_ucd_15pc}, the power law index is $\alpha = 1.98 \pm 0.02$. \cite{yang_flares} compiled a catalogue of flares using all available long cadence \textit{Kepler} light curves from DR25, and found an average $\alpha \approx 2$ power law index for all spectral types excluding A-stars, and also a $\sim 10$\% incidence rate of flare stars among M-dwarfs, consistent with our result ($\sim 8$\%). So it appears that the slope found for the TRAPPIST--1 analogues here ($\alpha = 2.11$) is typical for late M-dwarfs, and consistent with TRAPPIST--1 itself ($\alpha = 2.03$). It also confirms that the same power law index is further applicable for one or two orders of magnitude larger flares. According to \cite{aschwanden_alpha}, magnetic and thermal energies dominate around $\alpha \approx 2.0$, opposed to nonthermal energy around $\alpha \approx 1.4$.

\subsection{Habitability}

Out of a sample of 57 exoplanets studied by \cite{tardigrade_index}, TRAPPIST--1f is the closest to Earth in terms of Earth Similarity Index, Active and Cryptobiotic Tardigrade Index, meaning that life -- as we know it -- may survive on its surface (especially extremophile organisms like \textit{tardigrades}). And while \textit{Kepler} searched for exoplanets around solar-like stars, the number of detections around M-dwarfs is expected to grow with \textit{TESS}. So the question of habitability around UCDs -- especially TRAPPIST--1 -- is timely.

Late-type dwarfs are important for exoplanet habitability studies since they spend billions of years on the main sequence, giving life enough time to emerge on the surface of planets orbiting them. However, the magnetic activity of these stars can endanger the habitability of the orbiting planets. One of the hazards is the erosion of the planetary atmosphere through intense electromagnetic and particle radiation. To preserve their atmospheres, the planets need a strong magnetic field, even 10--1000 times stronger than Earth's magnetosphere \citep{vidotto_magnetosphere}. We know from the Sun that a large portion of flares are accompanied by CMEs, that are especially harmful if they reach the planetary atmosphere. 
Apart from the magnetosphere of the planets, \cite{t1-magnetic_field} argue that the host star itself can trap the CMEs if it has sufficient magnetic field. They calculate a global 1.4--1.75\,kG dipole field for TRAPPIST--1, which is strong enough to suppress even $10^{34}-10^{35}$\,erg events (assuming that flares are accompanied by CMEs with comparable energy). 
Furthermore, successful CMEs seem to be very rare in late-type dwarfs empirically, 90--98\% of the eruptions do not reach the escape velocity, suggesting that electromagnetic radiation would be a larger factor for atmosphere erosion than particles \citep{v374peg, CMEs}.
This suggests that the planets of TRAPPIST--1 are safe CME-wise assuming the FFD from Fig.~\ref{fig:FFD}. 
And even if energetic flares/CMEs would erode the atmosphere of young planets and deplete their water reservoir in the early stages of their evolution, the replenishment of the ocean is possible through asteroid bombardment events \citep{t1-waterdelivery-zoli}. 
In this scenario, late M-dwarfs are not likely to erode their secondary atmospheres, after their primordial H/He envelope is gone \citep{atmospheric_escape}. 
\cite{atmospheric_escape_t1} reached a similar conclusion for TRAPPIST--1, showing that the outer planets can retain their atmospheres for billions of years.

Once an atmosphere or even an ocean is present, abiogenesis is possible. The other hazardous effect of magnetic activity is the increased UV radiation during stellar flares that could harm simple or more evolved organisms (we note however that the UV radiation of flares might be the source that powers prebiotic photochemistry, see e.g. \citealt{flares-prebiotic}). \cite{ocean_habitability} studied the impact of superflares on two different bacteria (\textit{E. coli} and \textit{D. radiodurans}) on the TRAPPIST--1 planets. They found that an ozone layer or a few meters deep ocean is sufficient for the bacteria to survive the largest flares observed by \textit{Kepler} K2. \cite{proxima_uv} showed that ozone is not the only option for a protective atmosphere, and a combination of CO$_2$ and N$_2$ could also suffice.

\begin{figure}
    \centering
    \includegraphics[width=0.95\columnwidth]{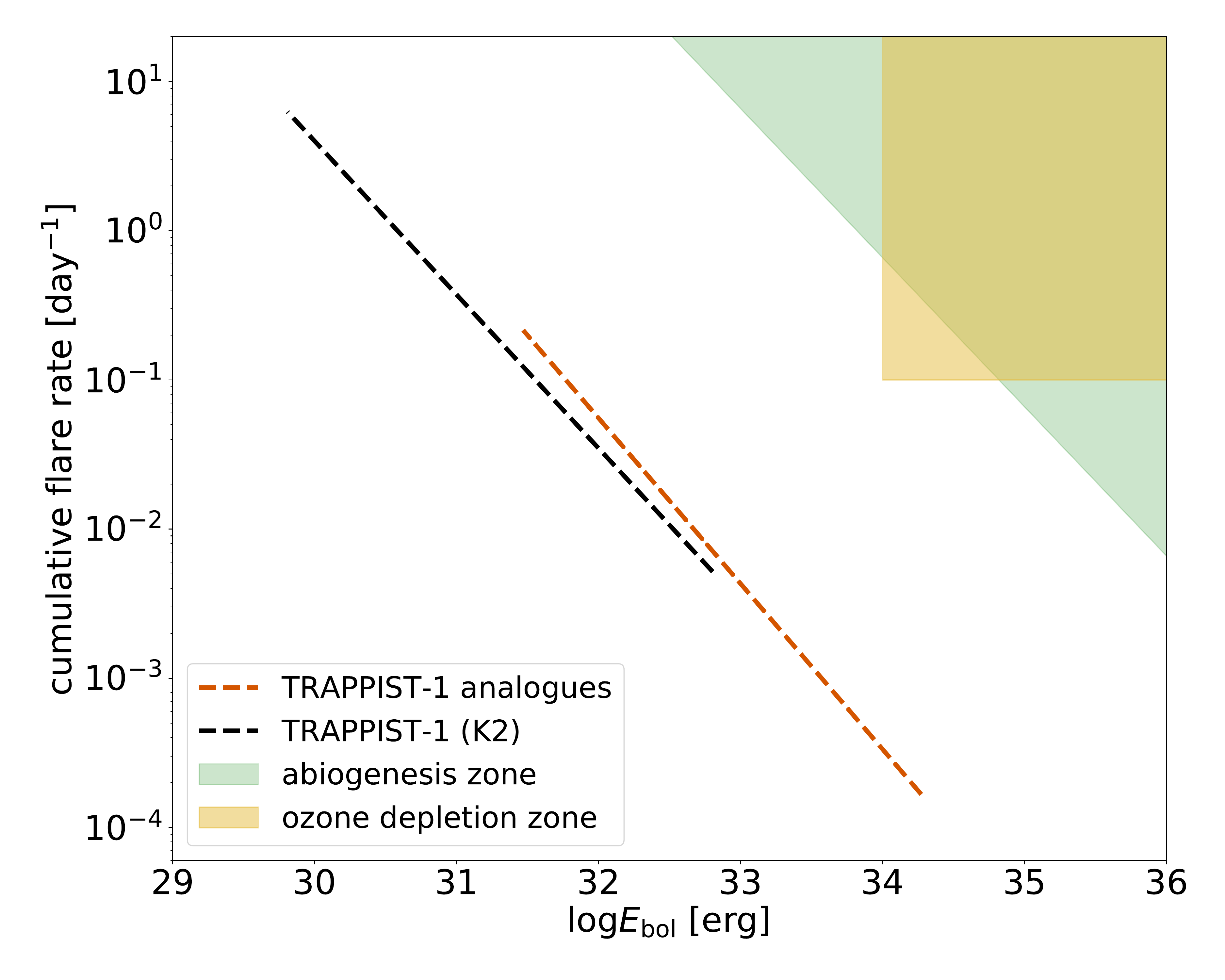}
    \caption{Flare frequency distribution showing the region where abiogenesis is possible \citep{abiogenesis}, and where ozone depletion can occur \citep{ozone_depletion}. Dashed lines are the linear fits from Fig.~\ref{fig:FFD}. We note the use of $E_\mathrm{bol}$ instead of $E_\mathrm{TESS}$.}
    \label{fig:FFD_habitability}
\end{figure}

The FFD presented here includes flares with higher energies than ever observed on TRAPPIST--1 before, and it seems that the FFD can still be described by a single power law with $\alpha \approx 2$. Figure \ref{fig:FFD_habitability} shows the schematic FFD along with regions where ozone depletion and abiogenesis can be expected. \cite{ozone_depletion} estimated that the ozone layer of unmagnetized planets can be fully destroyed if the rate of $E_\mathrm{bol} > 10^{34}$\,erg superflares reaches $0.1$\,day$^{-1}$. According to \cite{abiogenesis}, the lower limit on the FFD where abiogenesis is possible is also a power law, with $-1$ exponent. Their original relation used U-band energy, but we adopt the formula of \cite{evryscope} with $E_\mathrm{bol}$ (using blackbody conversion factors from \citealt{abiogenesis_conversion}), thus giving:
\begin{equation}
    \nu (E_\mathrm{bol}) > 6.6 \times 10^{33} \times E^{-1}_\mathrm{bol}
\end{equation}
And since $\alpha \approx 2$ corresponds to $1 - \alpha \approx -1$ exponent, the border of the abiogenesis zone is approximately parallel to the FFD of TRAPPIST--1, hinting that they do not intersect, not even at higher energies. These strengthen the findings of \cite{evryscope}, that the current flare rate of TRAPPIST--1 is unlikely to cause ozone depletion or initiate abiogenesis.

\section{Summary and conclusions}
\label{sec:conclusions}

We analysed 248 TRAPPIST--1 analogues using 30-min \textit{TESS} light curves, we can summarise our findings as follows:
\begin{enumerate}
    \item We found altogether 94 flare events on 21 stars. From the targets, in 42 a periodic light curve modulation was found (likely due to rotation) in the range of 0.1--5$^\mathrm{d}$ (we searched only for periods up to 5$^\mathrm{d}$). All the 21 stars with flares show rotational modulation as well.
    \item We estimated the approximate age of 88 stars from the sample with various methods, but did not find a convincing correlation between age, rotation rate and H$\alpha$ emission.
    \item The power law slope of the composite FFD was found to be similar to the value for TRAPPIST--1, hinting that it is not an especially active/inactive star for its spectral type. Combining the light curves of objects in the sample enabled us to find stronger flares than ever observed on TRAPPIST--1 before, suggesting that they actually occur, even if rarely. We found that flares with $E_\mathrm{TESS} > 10^{33}$\,erg ($E_\mathrm{bol} \gtrsim 10^{34}$\,erg) can be expected every few decades.
    \item Is seems that the flaring rate of TRAPPIST--1 is not sufficient to fully destroy the possible ozone layer of its planets, nor to initiate abiogenesis via UV radiation. \cite{evryscope} reached a similar conclusion, using an upper limit for the superflare rate from Evryscope observations.
\end{enumerate}

\begin{acknowledgements}
The authors would like to thank the anonymous referee for improving the quality of the paper with helpful comments and suggestions.

BS was supported by the \'UNKP-19-3 New National Excellence Program of the Ministry for Innovation and Technology.

KV was supported by the Bolyai J\'anos Research Scholarship of the Hungarian Academy of Sciences.

This project has been supported by the Lend\"ulet Program  of the Hungarian Academy of Sciences, project No. LP2018-7/2019, the 
NKFI KH-130526 and 
NKFI K-131508 grants, 
the Hungarian OTKA Grant No. 119993 and by the NKFI grant  2019-2.1.11-TÉT-2019-00056.

On behalf of \textit{"Analysis of space-borne photometric data"} project we thank for the usage of MTA Cloud (\url{https://cloud.mta.hu}) that helped us achieving the results published in this paper.

Authors acknowledge the financial support  of the Austrian-Hungarian  Action  Foundation (95 \"ou3, 98\"ou5, 101\"ou13). 

This paper includes data collected by the \textit{TESS} mission. Funding for the \textit{TESS} mission is provided by the NASA Explorer Program.

This research has benefitted from the Ultracool RIZzo Spectral Library (\url{http://dx.doi.org/10.5281/zenodo.11313}), maintained by Jonathan Gagn\'e and Kelle Cruz.
\end{acknowledgements}


\bibliographystyle{aa}
\bibliography{40098}

\begin{appendix}
\section{Measured rotational periods}
\begin{figure*}
    \centering
    \includegraphics[width=\textwidth]{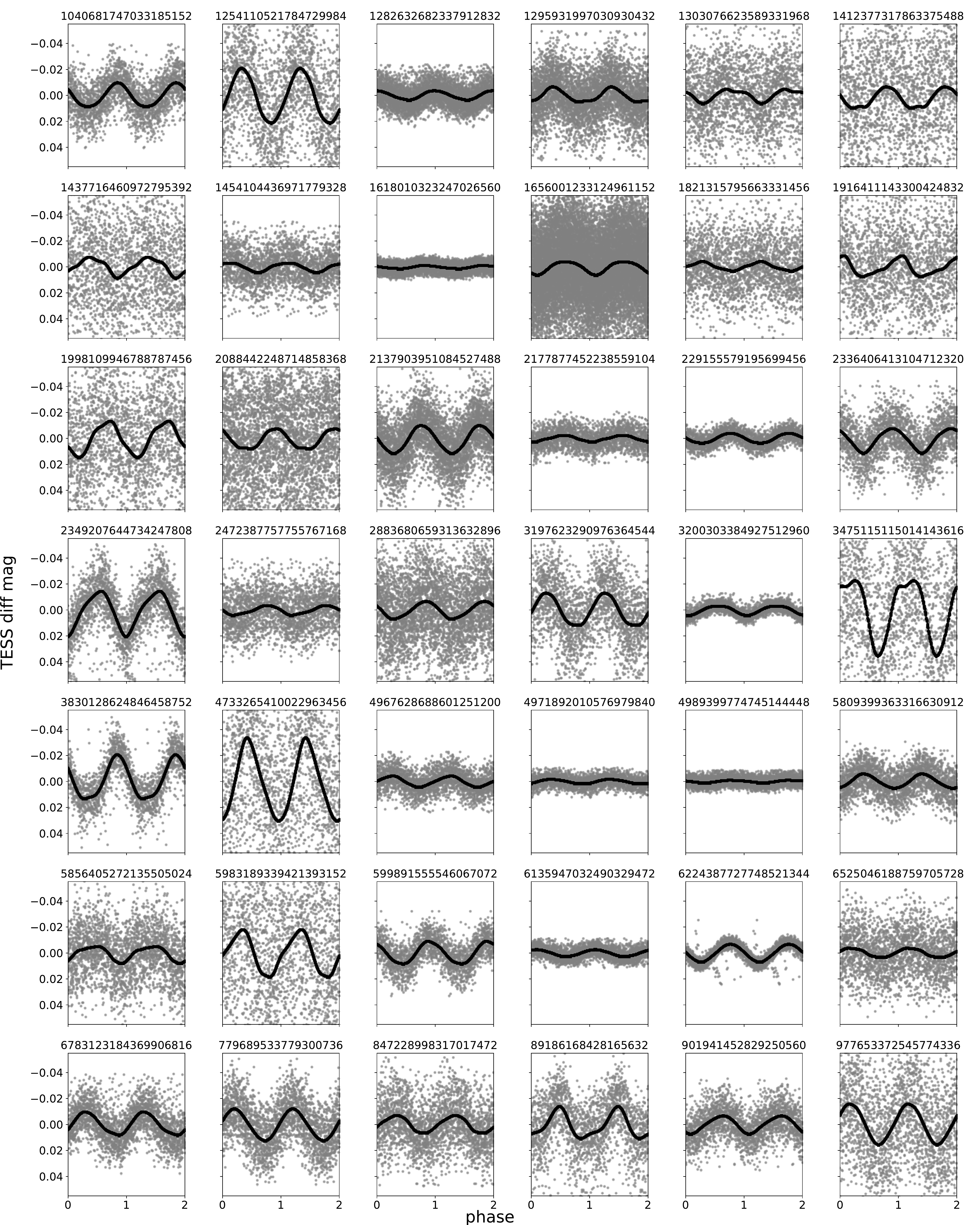}
    \caption{Light curves of stars with detected periodicity, folded with the average period from Table~\ref{table:periods}. Black line denotes the smoothed light curve.}
    \label{fig:all_periodic_phasecurves}
\end{figure*}

We present the rotational periods in Table~\ref{table:periods}. If the period was measurable in multiple sectors, the average period is given (the differences between the period values of the same star are consistent with the nominal error bars). In such cases the FAP value is calculated as the geometric mean of individual FAPs. The phase-folded light curves are plotted in Fig.~\ref{fig:all_periodic_phasecurves}.

\begin{table}
\caption{Rotational periods from the Lomb--Scargle analysis}\label{table:periods}
\centering
\begin{tabular}{lccccc}
\hline \hline
Gaia DR2 source ID & period & FAP & amplitude & scatter & \#flares \\
 & [days] &  & [mmag] & [mmag] &  \\
\hline
1040681747033185152 & $0.23882 \pm 0.00004$ & 3.5e-73 & 9.1 & 11.5 & 2 \\
1254110521784729984 & $0.26027 \pm 0.00009$ & 3.3e-78 & 21.1 & 32.0 & 0 \\
1282632682337912832 & $0.37110 \pm 0.00014$ & 8.9e-41 & 4.2 & 7.4 & 1 \\
1295931997030930432 & $0.25256 \pm 0.00011$ & 6.4e-22 & 6.3 & 15.5 & 5 \\
1303076623589331968 & $3.53028 \pm 0.04614$ & 1.7e-04 & 4.7 & 25.4 & 0 \\
1412377317863375488 & $0.91820 \pm 0.00370$ & 1.2e-05 & 10.3 & 78.1 & 0 \\
1437716460972795392 & $4.88480 \pm 0.09917$ & 2.9e-04 & 1.8 & 36.0 & 0 \\
1454104436971779328 & $0.64009 \pm 0.00092$ & 2.4e-14 & 3.6 & 12.1 & 1 \\
1618010323247026560 & $0.41250 \pm 0.00030$ & 2.2e-22 & 0.9 & 3.8 & 1 \\
1656001233124961152 & $0.81474 \pm 0.04714$ & 9.6e-05 & 4.8 & 25.3 & 3 \\
1821315795663331456 & $0.22591 \pm 0.05854$ & 2.0e-08 & 3.6 & 18.3 & 1 \\
1916411143300424832 & $0.82768 \pm 0.00189$ & 1.4e-08 & 9.4 & 28.2 & 0 \\
1998109946788787456 & $0.95529 \pm 0.00179$ & 1.9e-17 & 9.4 & 35.3 & 0 \\
2088442248714858368 & $1.45381 \pm 0.24790$ & 9.4e-11 & 9.4 & 39.7 & 0 \\
2137903951084527488 & $0.39450 \pm 0.00013$ & 2.6e-58 & 10.7 & 15.3 & 3 \\
2177877452238559104 & $0.35025 \pm 0.00022$ & 4.8e-19 & 2.7 & 6.6 & 0 \\
229155579195699456 & $0.98491 \pm 0.00101$ & 6.8e-95 & 3.7 & 4.7 & 0 \\
2336406413104712320 & $0.30911 \pm 0.00008$ & 2.9e-57 & 9.5 & 13.2 & 3 \\
2349207644734247808 & $2.13867 \pm 0.00221$ & 6.0e-161 & 20.9 & 13.6 & 1 \\
2472387757755767168 & $0.40031 \pm 0.00033$ & 5.1e-08 & 3.7 & 13.6 & 0 \\
2883680659313632896 & $1.92203 \pm 0.00778$ & 1.6e-08 & 6.8 & 23.4 & 1 \\
3197623290976364544 & $4.91369 \pm 0.02047$ & 2.0e-57 & 12.4 & 19.9 & 0 \\
3200303384927512960 & $0.50062 \pm 0.00019$ & 4.5e-124 & 3.8 & 3.3 & 10 \\
3475115115014143616 & $0.22498 \pm 0.00007$ & 3.5e-53 & 33.2 & 42.6 & 0 \\
3830128624846458752 & $0.21266 \pm 0.00003$ & 1.1e-208 & 16.9 & 10.0 & 3 \\
4733265410022963456 & $0.20460 \pm 0.00004$ & 2.0e-51 & 28.8 & 81.4 & 1 \\
4967628688601251200 & $0.40391 \pm 0.00020$ & 5.5e-46 & 4.2 & 7.0 & 1 \\
4971892010576979840 & $0.70254 \pm 0.00074$ & 7.1e-26 & 2.1 & 3.7 & 0 \\
4989399774745144448 & $1.99002 \pm 0.00607$ & 5.2e-13 & 0.9 & 2.8 & 5 \\
5809399363316630912 & $0.11082 \pm 0.00001$ & 4.9e-44 & 5.9 & 9.7 & 0 \\
5856405272135505024 & $0.12529 \pm 0.00006$ & 1.4e-19 & 7.0 & 17.9 & 0 \\
5983189339421393152 & $2.95347 \pm 0.01652$ & 3.0e-16 & 18.2 & 54.1 & 0 \\
599891555546067072 & $0.73416 \pm 0.00053$ & 4.8e-105 & 8.6 & 8.9 & 3 \\
6135947032490329472 & $1.47841 \pm 0.00148$ & 1.5e-119 & 2.6 & 2.9 & 3 \\
6224387727748521344 & $2.01107 \pm 0.00247$ & 1.7e-316 & 6.8 & 3.7 & 0 \\
6525046188759705728 & $0.15211 \pm 0.00008$ & 8.4e-05 & 3.9 & 16.3 & 1 \\
6783123184369906816 & $0.41521 \pm 0.00011$ & 2.2e-85 & 9.0 & 10.6 & 3 \\
779689533779300736 & $0.10527 \pm 0.00001$ & 8.8e-92 & 12.2 & 14.9 & 0 \\
847228998317017472 & $1.04848 \pm 0.00137$ & 6.6e-22 & 6.7 & 16.2 & 0 \\
89186168428165632 & $0.70691 \pm 0.00082$ & 4.0e-48 & 12.3 & 17.1 & 0 \\
901941452829250560 & $0.34495 \pm 0.00016$ & 2.4e-54 & 7.3 & 10.0 & 3 \\
977653372545774336 & $1.03019 \pm 0.00170$ & 2.1e-34 & 15.7 & 30.1 & 0 \\
\hline
\end{tabular}
\end{table}

\section{Flare events}
We present the parameters of all 94 individual flare events in Table~\ref{table:flares}. In some cases the MCMC fit did not converge, for those events $t_{1/2, \mathrm{fit}}$, $A_\mathrm{fit}$ and $ED_\mathrm{fit}$ are omitted, and $t_\mathrm{peak, fit}$ only indicates the time of the highest flux value.

\LTcapwidth=\textwidth 
\onecolumn

\begin{longtable}{lcccccc}
\caption{Flare parameters. The fitted values are removed where the MCMC fit did not converge. The column $E_\mathrm{sum}$ equals to $ED_\mathrm{sum}$ multiplied by the luminosity of TRAPPIST--1 in the \textit{TESS} bandpass.}\label{table:flares}\\
\hline\hline
Gaia DR2 source ID & $t_\mathrm{peak, fit}$ [JD-2~450~000] & $t_{1/2, \mathrm{fit}}$ [min] & $A_\mathrm{fit}$ & $ED_\mathrm{fit}$ [min] & $ED_\mathrm{sum}$ [min] & $E_\mathrm{sum} [10^{31} \mathrm{erg}]$\\
\hline
\endfirsthead
\caption{Flare parameters. The fitted values are removed where the MCMC fit did not converge. The column $E_\mathrm{sum}$ equals to $ED_\mathrm{sum}$ multiplied by the luminosity of TRAPPIST--1 in the \textit{TESS} bandpass (continued).}
\endhead
\endfoot
1040681747033185152 & $8856.732 \pm 0.010$ & $8.3 \pm 6.3$ & $0.28 \pm 0.31$ & $4.4 \pm 3.6$ & $3.6 \pm 1.5$ & $4.5 \pm 1.9$ \\
1040681747033185152 & $8864.857 \pm 0.007$ & $9.9 \pm 14.5$ & $0.12 \pm 0.16$ & $2.8 \pm 1.6$ & $1.6 \pm 1.2$ & $2.0 \pm 1.5$ \\
1078021436789455872 & $8891.802 \pm 0.001$ & $9.3 \pm 0.7$ & $1.67 \pm 0.17$ & $28.8 \pm 1.8$ & $55.6 \pm 1.3$ & $70.0 \pm 1.6$ \\
1269789664970320512 & $8974.403$ & -- & -- & -- & $2.3 \pm 2.4$ & $2.9 \pm 3.1$ \\
1282632682337912832 & $8972.577 \pm 0.000$ & $7.8 \pm 0.5$ & $2.98 \pm 0.27$ & $42.4 \pm 1.3$ & $23.7 \pm 1.0$ & $29.9 \pm 1.2$ \\
1295931997030930432 & $8946.826 \pm 0.000$ & $7.3 \pm 0.7$ & $2.80 \pm 0.36$ & $37.7 \pm 1.7$ & $32.5 \pm 1.2$ & $41.0 \pm 1.5$ \\
1295931997030930432 & $8935.294 \pm 0.196$ & $17.1 \pm 34.8$ & $0.05 \pm 0.09$ & $1.7 \pm 1.3$ & $1.7 \pm 1.0$ & $2.1 \pm 1.3$ \\
1295931997030930432 & $8949.608 \pm 0.006$ & $7.3 \pm 5.0$ & $0.23 \pm 0.23$ & $3.3 \pm 1.8$ & $1.6 \pm 1.0$ & $2.0 \pm 1.3$ \\
1295931997030930432 & $8950.715 \pm 0.007$ & $18.3 \pm 11.4$ & $0.12 \pm 0.09$ & $4.5 \pm 1.2$ & $3.9 \pm 1.0$ & $4.9 \pm 1.3$ \\
1295931997030930432 & $8967.940 \pm 0.007$ & $11.3 \pm 5.6$ & $0.32 \pm 0.21$ & $7.0 \pm 2.3$ & $5.2 \pm 1.3$ & $6.5 \pm 1.7$ \\
1324598017513898752 & $8979.130 \pm 0.006$ & $14.4 \pm 8.0$ & $0.28 \pm 0.22$ & $8.4 \pm 2.7$ & $6.8 \pm 1.9$ & $8.6 \pm 2.4$ \\
13533892921879936 & $8421.808 \pm 0.000$ & $11.2 \pm 0.8$ & $1.88 \pm 0.17$ & $38.6 \pm 1.4$ & $23.2 \pm 1.3$ & $29.2 \pm 1.7$ \\
13533892921879936 & $8426.255 \pm 0.006$ & $9.2 \pm 5.2$ & $0.26 \pm 0.22$ & $4.8 \pm 2.2$ & $4.3 \pm 1.2$ & $5.5 \pm 1.5$ \\
1454104436971779328 & $8953.733 \pm 0.007$ & $12.7 \pm 9.5$ & $0.11 \pm 0.11$ & $3.0 \pm 1.2$ & $2.3 \pm 0.9$ & $2.9 \pm 1.1$ \\
1516293918446402176 & $8932.112 \pm 0.068$ & $34.6 \pm 114.1$ & $0.10 \pm 0.12$ & $7.2 \pm 4.4$ & $5.6 \pm 2.7$ & $7.0 \pm 3.4$ \\
1595520568815311744 & $8959.654 \pm 0.004$ & $86.4 \pm 17.0$ & $0.05 \pm 0.01$ & $7.6 \pm 0.9$ & $4.8 \pm 0.5$ & $6.1 \pm 0.7$ \\
1618010323247026560 & $8742.805 \pm 0.000$ & $5.3 \pm 1.1$ & $0.32 \pm 0.11$ & $3.2 \pm 0.5$ & $1.3 \pm 0.2$ & $1.7 \pm 0.2$ \\
1656001233124961152 & $8859.091 \pm 0.005$ & $11.8 \pm 7.7$ & $0.32 \pm 0.22$ & $7.8 \pm 3.8$ & $6.9 \pm 3.1$ & $8.6 \pm 3.9$ \\
1656001233124961152 & $8905.739 \pm 0.002$ & $11.9 \pm 2.0$ & $1.74 \pm 0.38$ & $38.2 \pm 5.0$ & $51.9 \pm 2.9$ & $65.4 \pm 3.7$ \\
1656001233124961152 & $8971.909 \pm 0.003$ & $30.4 \pm 10.9$ & $0.42 \pm 0.15$ & $23.7 \pm 3.5$ & $19.8 \pm 3.0$ & $24.9 \pm 3.8$ \\
1688578285187648128 & $8881.346 \pm 0.000$ & $6.2 \pm 0.9$ & $1.53 \pm 0.33$ & $17.5 \pm 1.5$ & $14.1 \pm 0.8$ & $17.8 \pm 1.0$ \\
1688578285187648128 & $8887.909 \pm 0.000$ & $4.8 \pm 0.4$ & $3.67 \pm 0.63$ & $32.1 \pm 2.7$ & $10.1 \pm 0.9$ & $12.7 \pm 1.1$ \\
1688578285187648128 & $8896.180 \pm 0.000$ & $4.5 \pm 0.5$ & $2.77 \pm 0.63$ & $23.0 \pm 2.6$ & $7.1 \pm 0.9$ & $8.9 \pm 1.1$ \\
1821315795663331456 & $8684.947 \pm 0.003$ & $10.3 \pm 2.6$ & $1.48 \pm 0.54$ & $28.1 \pm 5.8$ & $36.9 \pm 3.3$ & $46.5 \pm 4.1$ \\
1902388491693623680 & $8759.986 \pm 0.007$ & $24.6 \pm 28.1$ & $0.10 \pm 0.11$ & $5.3 \pm 2.4$ & $4.2 \pm 1.9$ & $5.3 \pm 2.4$ \\
2137903951084527488 & $8683.944 \pm 0.006$ & $14.6 \pm 12.4$ & $0.17 \pm 0.16$ & $5.4 \pm 2.5$ & $4.0 \pm 2.3$ & $5.1 \pm 2.9$ \\
2137903951084527488 & $8694.827 \pm 0.000$ & $11.0 \pm 2.0$ & $2.12 \pm 0.45$ & $42.9 \pm 2.9$ & $31.6 \pm 2.6$ & $39.8 \pm 3.3$ \\
2137903951084527488 & $8699.179 \pm 0.166$ & $42.2 \pm 81.7$ & $0.08 \pm 0.10$ & $6.3 \pm 4.7$ & $6.1 \pm 3.0$ & $7.7 \pm 3.8$ \\
2267517023167123712 & $8892.860 \pm 0.007$ & $22.4 \pm 11.0$ & $0.23 \pm 0.14$ & $10.5 \pm 2.2$ & $7.8 \pm 1.4$ & $9.9 \pm 1.8$ \\
2331849006126794880 & $8373.005 \pm 0.006$ & $18.3 \pm 10.2$ & $0.04 \pm 0.03$ & $1.7 \pm 0.5$ & $1.5 \pm 0.4$ & $1.9 \pm 0.5$ \\
2331849006126794880 & $8379.617 \pm 0.008$ & $8.5 \pm 4.6$ & $0.18 \pm 0.13$ & $2.8 \pm 0.8$ & $1.8 \pm 0.4$ & $2.3 \pm 0.5$ \\
2336406413104712320 & $8361.626 \pm 0.000$ & $15.2 \pm 0.2$ & $12.04 \pm 0.28$ & $335.5 \pm 3.6$ & $170.8 \pm 2.9$ & $215.2 \pm 3.6$ \\
2336406413104712320 & $8369.367 \pm 0.000$ & $33.7 \pm 1.1$ & $2.32 \pm 0.05$ & $143.1 \pm 1.8$ & $163.0 \pm 2.5$ & $205.3 \pm 3.1$ \\
2336406413104712320 & $8373.175 \pm 0.003$ & $6.4 \pm 2.1$ & $0.48 \pm 0.35$ & $5.9 \pm 2.7$ & $7.7 \pm 1.3$ & $9.7 \pm 1.6$ \\
2349207644734247808 & $8396.660 \pm 0.000$ & $6.9 \pm 0.1$ & $16.41 \pm 0.43$ & $206.1 \pm 2.0$ & $247.8 \pm 2.8$ & $312.2 \pm 3.5$ \\
2523162448812944640 & $8404.447 \pm 0.003$ & $13.8 \pm 4.0$ & $0.76 \pm 0.27$ & $20.3 \pm 4.6$ & $23.5 \pm 3.0$ & $29.6 \pm 3.8$ \\
2523162448812944640 & $8405.779 \pm 0.086$ & $34.9 \pm 73.7$ & $0.11 \pm 0.12$ & $7.5 \pm 4.9$ & $6.1 \pm 3.1$ & $7.6 \pm 4.0$ \\
2883680659313632896 & $8482.405 \pm 0.003$ & $11.3 \pm 4.3$ & $0.57 \pm 0.24$ & $12.7 \pm 4.2$ & $15.4 \pm 2.8$ & $19.4 \pm 3.5$ \\
3200303384927512960 & $8440.910$ & -- & -- & -- & $1.5 \pm 0.2$ & $1.8 \pm 0.3$ \\
3200303384927512960 & $8442.219$ & -- & -- & -- & $6.4 \pm 0.2$ & $8.1 \pm 0.3$ \\
3200303384927512960 & $8446.240$ & -- & -- & -- & $4.4 \pm 0.2$ & $5.6 \pm 0.3$ \\
3200303384927512960 & $8447.136$ & -- & -- & -- & $2.0 \pm 0.2$ & $2.5 \pm 0.2$ \\
3200303384927512960 & $8451.596 \pm 0.000$ & $6.6 \pm 0.3$ & $1.52 \pm 0.09$ & $18.4 \pm 0.4$ & $19.3 \pm 0.1$ & $24.3 \pm 0.2$ \\
3200303384927512960 & $8452.846 \pm 0.000$ & $5.7 \pm 0.0$ & $10.57 \pm 0.11$ & $111.0 \pm 0.5$ & $90.8 \pm 0.2$ & $114.4 \pm 0.2$ \\
3200303384927512960 & $8455.366$ & -- & -- & -- & $1.7 \pm 0.2$ & $2.2 \pm 0.3$ \\
3200303384927512960 & $8458.530$ & -- & -- & -- & $1.7 \pm 0.2$ & $2.2 \pm 0.3$ \\
3200303384927512960 & $8458.847 \pm 0.000$ & $4.9 \pm 0.2$ & $2.57 \pm 0.18$ & $23.0 \pm 0.7$ & $11.6 \pm 0.2$ & $14.7 \pm 0.3$ \\
3200303384927512960 & $8459.897 \pm 0.006$ & $6.0 \pm 3.8$ & $0.09 \pm 0.09$ & $1.0 \pm 0.5$ & $0.4 \pm 0.2$ & $0.5 \pm 0.3$ \\
3830128624846458752 & $8520.216 \pm 0.004$ & $6.8 \pm 2.1$ & $0.51 \pm 0.42$ & $6.6 \pm 3.2$ & $8.1 \pm 0.9$ & $10.2 \pm 1.1$ \\
3830128624846458752 & $8521.045 \pm 0.006$ & $12.5 \pm 7.5$ & $0.20 \pm 0.17$ & $5.4 \pm 2.2$ & $3.7 \pm 0.8$ & $4.7 \pm 1.1$ \\
3830128624846458752 & $8527.493 \pm 0.000$ & $10.0 \pm 1.3$ & $0.91 \pm 0.14$ & $16.6 \pm 0.7$ & $15.5 \pm 0.6$ & $19.6 \pm 0.8$ \\
4694260956582093440 & $8360.236$ & -- & -- & -- & $5.3 \pm 3.3$ & $6.6 \pm 4.2$ \\
4733265410022963456 & $8393.489 \pm 0.001$ & $12.1 \pm 2.4$ & $0.89 \pm 0.15$ & $20.6 \pm 3.7$ & $30.9 \pm 2.9$ & $38.9 \pm 3.7$ \\
4742124385662135168 & $8360.320 \pm 0.006$ & $18.1 \pm 9.8$ & $0.26 \pm 0.17$ & $9.7 \pm 2.9$ & $7.7 \pm 2.4$ & $9.7 \pm 3.1$ \\
4825880783419986432 & $8458.730 \pm 0.006$ & $8.8 \pm 2.3$ & $0.43 \pm 0.20$ & $6.9 \pm 2.3$ & $4.5 \pm 0.7$ & $5.7 \pm 0.9$ \\
4866773609425146496 & $8442.740$ & -- & -- & -- & $2.0 \pm 2.5$ & $2.5 \pm 3.1$ \\
4967628688601251200 & $8400.533 \pm 0.008$ & $19.0 \pm 15.0$ & $0.05 \pm 0.05$ & $2.0 \pm 0.6$ & $1.4 \pm 0.6$ & $1.7 \pm 0.8$ \\
4989399774745144448 & $8365.200 \pm 0.000$ & $4.1 \pm 0.6$ & $1.14 \pm 0.39$ & $8.4 \pm 1.5$ & $5.8 \pm 0.2$ & $7.3 \pm 0.3$ \\
4989399774745144448 & $8371.037 \pm 0.002$ & $11.3 \pm 7.7$ & $0.09 \pm 0.06$ & $1.8 \pm 0.3$ & $1.1 \pm 0.3$ & $1.4 \pm 0.3$ \\
4989399774745144448 & $8389.263 \pm 0.000$ & $4.2 \pm 1.1$ & $0.45 \pm 0.26$ & $3.4 \pm 1.0$ & $1.2 \pm 0.2$ & $1.5 \pm 0.3$ \\
4989399774745144448 & $8399.930 \pm 0.000$ & $5.1 \pm 1.6$ & $0.23 \pm 0.13$ & $2.2 \pm 0.5$ & $1.5 \pm 0.2$ & $1.9 \pm 0.3$ \\
4989399774745144448 & $8400.896 \pm 0.006$ & $4.4 \pm 1.8$ & $0.18 \pm 0.14$ & $1.4 \pm 0.8$ & $0.6 \pm 0.2$ & $0.7 \pm 0.3$ \\
5055805741577757824 & $8414.170 \pm 0.006$ & $3.9 \pm 2.3$ & $0.21 \pm 0.20$ & $1.5 \pm 1.1$ & $0.9 \pm 0.3$ & $1.2 \pm 0.4$ \\
5055805741577757824 & $8418.171 \pm 0.229$ & $12.2 \pm 34.2$ & $0.05 \pm 0.12$ & $0.9 \pm 3.5$ & $0.4 \pm 0.3$ & $0.5 \pm 0.4$ \\
5055805741577757824 & $8429.918 \pm 0.006$ & $5.9 \pm 4.9$ & $0.09 \pm 0.10$ & $1.0 \pm 0.6$ & $0.4 \pm 0.3$ & $0.5 \pm 0.3$ \\
5433620854830342272 & $8545.027 \pm 0.006$ & $15.4 \pm 9.4$ & $0.30 \pm 0.22$ & $9.6 \pm 3.5$ & $8.6 \pm 2.6$ & $10.8 \pm 3.3$ \\
5563853506009853568 & $8483.796 \pm 0.006$ & $9.3 \pm 3.9$ & $0.46 \pm 0.30$ & $7.9 \pm 3.2$ & $6.4 \pm 1.3$ & $8.1 \pm 1.6$ \\
5637175400984142336 & $8522.027 \pm 0.009$ & $25.4 \pm 28.7$ & $0.11 \pm 0.11$ & $5.6 \pm 2.6$ & $4.3 \pm 2.1$ & $5.4 \pm 2.6$ \\
5661194163772723072 & $8525.277 \pm 0.005$ & $4.9 \pm 2.4$ & $0.29 \pm 0.30$ & $2.6 \pm 2.0$ & $2.6 \pm 0.6$ & $3.2 \pm 0.8$ \\
5661194163772723072 & $8563.632 \pm 0.005$ & $14.3 \pm 11.5$ & $0.12 \pm 0.13$ & $4.0 \pm 1.7$ & $2.7 \pm 1.1$ & $3.4 \pm 1.4$ \\
5884382654716110848 & $8631.969$ & -- & -- & -- & $1.2 \pm 1.0$ & $1.6 \pm 1.2$ \\
5884382654716110848 & $8649.319$ & -- & -- & -- & $0.6 \pm 0.5$ & $0.8 \pm 0.7$ \\
5891504088488622208 & $8607.471 \pm 0.009$ & $31.0 \pm 27.6$ & $0.15 \pm 0.13$ & $9.3 \pm 3.8$ & $7.0 \pm 3.4$ & $8.8 \pm 4.2$ \\
5891504088488622208 & $8609.402 \pm 0.006$ & $15.2 \pm 7.8$ & $0.31 \pm 0.20$ & $9.6 \pm 3.2$ & $7.3 \pm 2.3$ & $9.3 \pm 2.9$ \\
599891555546067072 & $8497.566 \pm 0.006$ & $7.4 \pm 5.6$ & $0.16 \pm 0.18$ & $2.4 \pm 1.3$ & $1.4 \pm 0.8$ & $1.7 \pm 1.0$ \\
599891555546067072 & $8499.771 \pm 0.006$ & $19.3 \pm 19.2$ & $0.09 \pm 0.10$ & $3.8 \pm 1.2$ & $2.5 \pm 0.7$ & $3.2 \pm 0.9$ \\
599891555546067072 & $8509.263 \pm 0.000$ & $4.8 \pm 0.8$ & $1.12 \pm 0.36$ & $9.8 \pm 1.6$ & $3.8 \pm 0.6$ & $4.8 \pm 0.7$ \\
6135947032490329472 & $8590.261 \pm 0.000$ & $8.2 \pm 0.2$ & $2.39 \pm 0.05$ & $35.6 \pm 0.2$ & $76.1 \pm 0.2$ & $95.9 \pm 0.3$ \\
6135947032490329472 & $8615.856 \pm 0.006$ & $3.8 \pm 1.5$ & $0.32 \pm 0.27$ & $2.1 \pm 1.5$ & $1.1 \pm 0.2$ & $1.4 \pm 0.3$ \\
6135947032490329472 & $8622.737 \pm 0.009$ & $16.0 \pm 20.2$ & $0.02 \pm 0.03$ & $0.7 \pm 0.3$ & $0.5 \pm 0.3$ & $0.7 \pm 0.3$ \\
6525046188759705728 & $8376.547 \pm 0.007$ & $17.6 \pm 14.7$ & $0.14 \pm 0.14$ & $5.5 \pm 2.0$ & $4.1 \pm 1.7$ & $5.2 \pm 2.2$ \\
6783123184369906816 & $8327.900 \pm 0.006$ & $7.8 \pm 5.5$ & $0.20 \pm 0.20$ & $3.2 \pm 1.7$ & $1.4 \pm 1.3$ & $1.7 \pm 1.6$ \\
6783123184369906816 & $8345.119 \pm 0.001$ & $9.7 \pm 7.3$ & $0.33 \pm 0.26$ & $6.2 \pm 1.2$ & $4.1 \pm 1.0$ & $5.2 \pm 1.2$ \\
6783123184369906816 & $8345.118 \pm 0.001$ & $8.7 \pm 5.9$ & $0.39 \pm 0.27$ & $6.4 \pm 1.3$ & $4.1 \pm 0.9$ & $5.2 \pm 1.2$ \\
697274883805771776 & $8878.651 \pm 0.005$ & $10.8 \pm 1.6$ & $2.83 \pm 0.91$ & $56.4 \pm 14.0$ & $56.5 \pm 2.2$ & $71.2 \pm 2.8$ \\
706729107552444672 & $8873.242 \pm 0.005$ & $31.2 \pm 15.0$ & $0.22 \pm 0.10$ & $12.9 \pm 2.9$ & $11.4 \pm 2.4$ & $14.4 \pm 3.0$ \\
901941452829250560 & $8846.097 \pm 0.000$ & $6.2 \pm 1.0$ & $1.45 \pm 0.36$ & $16.5 \pm 1.7$ & $7.2 \pm 1.0$ & $9.1 \pm 1.2$ \\
901941452829250560 & $8860.962 \pm 0.006$ & $8.8 \pm 3.6$ & $0.36 \pm 0.23$ & $5.8 \pm 2.4$ & $4.2 \pm 0.9$ & $5.3 \pm 1.1$ \\
901941452829250560 & $8865.911 \pm 0.216$ & $36.0 \pm 67.2$ & $0.04 \pm 0.06$ & $2.9 \pm 1.9$ & $3.0 \pm 1.3$ & $3.8 \pm 1.6$ \\
947463811200288512 & $8852.468 \pm 0.001$ & $5.9 \pm 0.7$ & $1.85 \pm 0.38$ & $20.0 \pm 2.5$ & $52.7 \pm 1.4$ & $66.4 \pm 1.7$ \\
947463811200288512 & $8867.782$ & -- & -- & -- & $5.5 \pm 5.5$ & $6.9 \pm 6.9$ \\
1123855296251645824 & $9031.140 \pm 0.001$ & $16.3 \pm 9.2$ & $0.45 \pm 0.29$ & $14.1 \pm 2.3$ & $10.2 \pm 2.1$ & $12.8 \pm 2.7$ \\
1125848440251239424 & $9033.948$ & -- & -- & -- & $7.4 \pm 4.7$ & $9.3 \pm 5.9$ \\
1361972479325240960 & $9024.755 \pm 0.007$ & $18.2 \pm 7.3$ & $0.50 \pm 0.30$ & $18.0 \pm 7.0$ & $19.7 \pm 3.4$ & $24.8 \pm 4.3$ \\
\hline
\end{longtable}

\section{Comoving pairs}\label{appendix:comoving_pairs}
We present the parameters of 19 comoving pair candidates in Table~\ref{table:comoving_pairs}.

\begin{table}[b]
\caption{Comoving pair candidates from \textit{Gaia} DR2}\label{table:comoving_pairs}
\centering
\begin{tabular}{ccccc}
\hline \hline
target & companion & separation & kinematic age & comment \\
Gaia DR2 source ID & \textit{Gaia} DR2 source ID & [au] & [Gyr] & \\
\hline
   779689533779300736 &  779689606794219136 & 390 & - & M6.5Ve \\
   794031395948224640 &  794031395950867968 & 240 & - & \\
  1412377317863375488 & 1412377493957462784 & 2060 & $3.7 \pm 3.3$ & K3V with $v_{\mathrm{rad}}$ \\
  1688578285187648128 & 1688578280891900416 & 180 & - & M3.5V \\
  2088442248714858368 & 2088442248714858624 & 90 & - & \\
  2117179153332367232 & 2117179153330391040 & 830 & - & brown dwarf candidate \\
  2310240437249834496 & 2310240372825841280 & 750 & - & \\
  3732207661169873792 & 3732206772112073984 & 6190 & $6.1 \pm 3.4$ & K2 with $v_{\mathrm{rad}}$ \\
  3999432971480096640 & 3999433452516434560 & 5050 & $5.6 \pm 3.0$ & M3V with $v_{\mathrm{rad}}$ \\
  4126080699371400064 & 4126080699385397888 & 440 & - & \\
  4293315765165489536 & 4293318823182081408 & 440 & $4.0 \pm 3.3$ & M3V with $v_{\mathrm{rad}}$ \\
  4339417394313320192 & 4339465360508118912 & 1940 & $5.4 \pm 3.3$ & M3.5V with $v_{\mathrm{rad}}$ \\
  5070595856596619136 & 5070595856597335040 & 510 & $1.8 \pm 3.0$ & M with $v_{\mathrm{rad}}$ \\
  5627865354993448960 & 5627865694299883008 & 9920 & $1.6 \pm 3.0$ & G9V with $v_{\mathrm{rad}}$ \\
  5627865354993448960 & 5627865698590777728 & 9970 & - & \\
  5809399363316630912 & 5809399363316632064 & 70 & - & \\
  6357834388848708224 & 6357835694518769408 & 6220 & $0.8 \pm 2.6$ & G0V with $v_{\mathrm{rad}}$ \\
  6503514658710993664 & 6503514697366717568 & 1810 & - & M4 \\
  6645843007247800704 & 6645843007249648768 & 780 & $5.8 \pm 3.4$ & M2 with $v_{\mathrm{rad}}$ \\
  6783123184369906816 & 6783123184369906944 & 60 & - & \\
\hline
\end{tabular}
\end{table}

\end{appendix}

\end{document}